# The Pierre Auger Observatory I: The Cosmic Ray Energy Spectrum and Related Measurements

THE PIERRE AUGER COLLABORATION

*Observatorio Pierre Auger, Av. San Martín Norte 304, 5613 Malargüe, Argentina*

# The Pierre Auger Collaboration

P. Abreu[74], M. Aglietta[57], E.J. Ahn[93], I.F.M. Albuquerque[19], D. Allard[33], I. Allekotte[1], J. Allen[96], P. Allison[98], J. Alvarez Castillo[67], J. Alvarez-Muñiz[84], M. Ambrosio[50], A. Aminaei[68], L. Anchordoqui[109], S. Andringa[74], T. Antičić[27], A. Anzalone[56], C. Aramo[50], E. Arganda[81], F. Arqueros[81], H. Asorey[1], P. Assis[74], J. Aublin[35], M. Ave[41], M. Avenier[36], G. Avila[12], T. Bäcker[45], M. Balzer[40], K.B. Barber[13], A.F. Barbosa[16], R. Bardenet[34], S.L.C. Barroso[22], B. Baughman[98], J. Bäuml[39], J.J. Beatty[98], B.R. Becker[106], K.H. Becker[38], A. Bellétoile[37], J.A. Bellido[13], S. BenZvi[108], C. Berat[36], X. Bertou[1], P.L. Biermann[42], P. Billoir[35], F. Blanco[81], M. Blanco[82], C. Bleve[38], H. Blümer[41, 39], M. Boháčová[29, 101], D. Boncioli[51], C. Bonifazi[25, 35], R. Bonino[57], N. Borodai[72], J. Brack[91], P. Brogueira[74], W.C. Brown[92], R. Bruijn[87], P. Buchholz[45], A. Bueno[83], R.E. Burton[89], K.S. Caballero-Mora[99], L. Caramete[42], R. Caruso[52], A. Castellina[57], O. Catalano[56], G. Cataldi[49], L. Cazon[74], R. Cester[53], J. Chauvin[36], S.H. Cheng[99], A. Chiavassa[57], J.A. Chinellato[20], A. Chou[93, 96], J. Chudoba[29], R.W. Clay[13], M.R. Coluccia[49], R. Conceição[74], F. Contreras[11], H. Cook[87], M.J. Cooper[13], J. Coppens[68, 70], A. Cordier[34], U. Cotti[66], S. Coutu[99], C.E. Covault[89], A. Creusot[33, 79], A. Criss[99], J. Cronin[101], A. Curutiu[42], S. Dagoret-Campagne[34], R. Dallier[37], S. Dasso[8, 4], K. Daumiller[39], B.R. Dawson[13], R.M. de Almeida[26], M. De Domenico[52], C. De Donato[67, 48], S.J. de Jong[68, 70], G. De La Vega[10], W.J.M. de Mello Junior[20], J.R.T. de Mello Neto[25], I. De Mitri[49], V. de Souza[18], K.D. de Vries[69], G. Decerprit[33], L. del Peral[82], O. Deligny[32], H. Dembinski[41], N. Dhital[95], C. Di Giulio[47, 51], J.C. Diaz[95], M.L. Díaz Castro[17], P.N. Diep[110], C. Dobrigkeit [20], W. Docters[69], J.C. D'Olivo[67], P.N. Dong[110, 32], A. Dorofeev[91], J.C. dos Anjos[16], M.T. Dova[7], D. D'Urso[50], I. Dutan[42], J. Ebr[29], R. Engel[39], M. Erdmann[43], C.O. Escobar[20], A. Etchegoyen[2], P. Facal San Luis[101], I. Fajardo Tapia[67], H. Falcke[68, 71], G. Farrar[96], A.C. Fauth[20], N. Fazzini[93], A.P. Ferguson[89], A. Ferrero[2], B. Fick[95], A. Filevich[2], A. Filipčič[78, 79], S. Fliescher[43], C.E. Fracchiolla[91], E.D. Fraenkel[69], U. Fröhlich[45], B. Fuchs[16], R. Gaior[35], R.F. Gamarra[2], S. Gambetta[46], B. García[10], D. García Gámez[83], D. Garcia-Pinto[81], A. Gascon[83], H. Gemmeke[40], K. Gesterling[106], P.L. Ghia[35, 57], U. Giaccari[49], M. Giller[73], H. Glass[93], M.S. Gold[106], G. Golup[1], F. Gomez Albarracin[7], M. Gómez Berisso[1], P. Gonçalves[74], D. Gonzalez[41], J.G. Gonzalez[41], B. Gookin[91], D. Góra[41, 72], A. Gorgi[57], P. Gouffon[19], S.R. Gozzini[87], E. Grashorn[98], S. Grebe[68, 70], N. Griffith[98], M. Grigat[43], A.F. Grillo[58], Y. Guardincerri[4], F. Guarino[50], G.P. Guedes[21], A. Guzman[67], J.D. Hague[106], P. Hansen[7], D. Harari[1], S. Harmsma[69, 70], J.L. Harton[91], A. Haungs[39], T. Hebbeker[43], D. Heck[39], A.E. Herve[13], C. Hojvat[93], N. Hollon[101], V.C. Holmes[13], P. Homola[72], J.R. Hörandel[68], A. Horneffer[68], M. Hrabovský[30, 29], T. Huege[39], A. Insolia[52], F. Ionita[101], A. Italiano[52], C. Jarne[7], S. Jiraskova[68], M. Josebachuili[2], K. Kadija[27], K.-H. Kampert[38], P. Karhan[28], P. Kasper[93], B. Kégl[34], B. Keilhauer[39], A. Keivani[94], J.L. Kelley[68], E. Kemp[20], R.M. Kieckhafer[95], H.O. Klages[39], M. Kleifges[40], J. Kleinfeller[39], J. Knapp[87], D.-H. Koang[36], K. Kotera[101], N. Krohm[38], O. Krömer[40], D. Kruppke-Hansen[38], F. Kuehn[93], D. Kuempel[38], J.K. Kulbartz[44], N. Kunka[40], G. La Rosa[56], C. Lachaud[33], P. Lautridou[37], M.S.A.B. Leão[24], D. Lebrun[36], P. Lebrun[93], M.A. Leigui de Oliveira[24], A. Lemiere[32], A. Letessier-Selvon[35], I. Lhenry-Yvon[32], K. Link[41], R. López[63], A. Lopez Agüera[84], K. Louedec[34], J. Lozano Bahilo[83], A. Lucero[2, 57], M. Ludwig[41], H. Lyberis[32], M.C. Maccarone[56], C. Macolino[35], S. Maldera[57], D. Mandat[29], P. Mantsch[93], A.G. Mariazzi[7], J. Marin[11, 57], V. Marin[37], I.C. Maris[35], H.R. Marquez Falcon[66], G. Marsella[54], D. Martello[49], L. Martin[37], H. Martinez[64], O. Martínez Bravo[63],

H.J. Mathes[39], J. Matthews[94, 100], J.A.J. Matthews[106], G. Matthiae[51], D. Maurizio[53], P.O. Mazur[93], G. Medina-Tanco[67], M. Melissas[41], D. Melo[2, 53], E. Menichetti[53], A. Menshikov[40], P. Mertsch[85], C. Meurer[43], S. Mićanović[27], M.I. Micheletti[9], W. Miller[106], L. Miramonti[48], S. Mollerach[1], M. Monasor[101], D. Monnier Ragaigne[34], F. Montanet[36], B. Morales[67], C. Morello[57], E. Moreno[63], J.C. Moreno[7], C. Morris[98], M. Mostafá[91], C.A. Moura[24, 50], S. Mueller[39], M.A. Muller[20], G. Müller[43], M. Münchmeyer[35], R. Mussa[53], G. Navarra[57 †], J.L. Navarro[83], S. Navas[83], P. Necesal[29], L. Nellen[67], A. Nelles[68, 70], J. Neuser[38], P.T. Nhung[110], L. Niemietz[38], N. Nierstenhoefer[38], D. Nitz[95], D. Nosek[28], L. Nožka[29], M. Nyklicek[29], J. Oehlschläger[39], A. Olinto[101], V.M. Olmos-Gilbaja[84], M. Ortiz[81], N. Pacheco[82], D. Pakk Selmi-Dei[20], M. Palatka[29], J. Pallotta[3], N. Palmieri[41], G. Parente[84], E. Parizot[33], A. Parra[84], R.D. Parsons[87], S. Pastor[80], T. Paul[97], M. Pech[29], J. Pękala[72], R. Pelayo[84], I.M. Pepe[23], L. Perrone[54], R. Pesce[46], E. Petermann[105], S. Petrera[47], P. Petrinca[51], A. Petrolini[46], Y. Petrov[91], J. Petrovic[70], C. Pfendner[108], N. Phan[106], R. Piegaia[4], T. Pierog[39], P. Pieroni[4], M. Pimenta[74], V. Pirronello[52], M. Platino[2], V.H. Ponce[1], M. Pontz[45], P. Privitera[101], M. Prouza[29], E.J. Quel[3], S. Querchfeld[38], J. Rautenberg[38], O. Ravel[37], D. Ravignani[2], B. Revenu[37], J. Ridky[29], S. Riggi[84, 52], M. Risse[45], P. Ristori[3], H. Rivera[48], V. Rizi[47], J. Roberts[96], C. Robledo[63], W. Rodrigues de Carvalho[84, 19], G. Rodriguez[84], J. Rodriguez Martino[11, 52], J. Rodriguez Rojo[11], I. Rodriguez-Cabo[84], M.D. Rodríguez-Frías[82], G. Ros[82], J. Rosado[81], T. Rossler[30], M. Roth[39], B. Rouillé-d'Orfeuil[101], E. Roulet[1], A.C. Rovero[8], C. Rühle[40], F. Salamida[47, 39], H. Salazar[63], G. Salina[51], F. Sánchez[2], M. Santander[11], C.E. Santo[74], E. Santos[74], E.M. Santos[25], F. Sarazin[90], B. Sarkar[38], S. Sarkar[85], R. Sato[11], N. Scharf[43], V. Scherini[48], H. Schieler[39], P. Schiffer[43], A. Schmidt[40], F. Schmidt[101], O. Scholten[69], H. Schoorlemmer[68, 70], J. Schovancova[29], P. Schovánek[29], F. Schröder[39], S. Schulte[43], D. Schuster[90], S.J. Sciutto[7], M. Scuderi[52], A. Segreto[56], M. Settimo[45], A. Shadkam[94], R.C. Shellard[16, 17], I. Sidelnik[2], G. Sigl[44], H.H. Silva Lopez[67], A. Śmiałkowski[73], R. Šmída[39, 29], G.R. Snow[105], P. Sommers[99], J. Sorokin[13], H. Spinka[88, 93], R. Squartini[11], S. Stanic[79], J. Stapleton[98], J. Stasielak[72], M. Stephan[43], E. Strazzeri[56], A. Stutz[36], F. Suarez[2], T. Suomijärvi[32], A.D. Supanitsky[8, 67], T. Šuša[27], M.S. Sutherland[94, 98], J. Swain[97], Z. Szadkowski[73], M. Szuba[39], A. Tamashiro[8], A. Tapia[2], M. Tartare[36], O. Taşcău[38], C.G. Tavera Ruiz[67], R. Tcaciuc[45], D. Tegolo[52, 61], N.T. Thao[110], D. Thomas[91], J. Tiffenberg[4], C. Timmermans[70, 68], D.K. Tiwari[66], W. Tkaczyk[73], C.J. Todero Peixoto[18, 24], B. Tomé[74], A. Tonachini[53], P. Travnicek[29], D.B. Tridapalli[19], G. Tristram[33], E. Trovato[52], M. Tueros[84, 4], R. Ulrich[99, 39], M. Unger[39], M. Urban[34], J.F. Valdés Galicia[67], I. Valiño[84, 39], L. Valore[50], A.M. van den Berg[69], E. Varela[63], B. Vargas Cárdenas[67], J.R. Vázquez[81], R.A. Vázquez[84], D. Veberič[79, 78], V. Verzi[51], J. Vicha[29], M. Videla[10], L. Villaseñor[66], H. Wahlberg[7], P. Wahrlich[13], O. Wainberg[2], D. Walz[43], D. Warner[91], A.A. Watson[87], M. Weber[40], K. Weidenhaupt[43], A. Weindl[39], S. Westerhoff[108], B.J. Whelan[13], G. Wieczorek[73], L. Wiencke[90], B. Wilczyńska[72], H. Wilczyński[72], M. Will[39], C. Williams[101], T. Winchen[43], L. Winders[109], M.G. Winnick[13], M. Wommer[39], B. Wundheiler[2], T. Yamamoto[101 a], T. Yapici[95], P. Younk[45], G. Yuan[94], A. Yushkov[84, 50], B. Zamorano[83], E. Zas[84], D. Zavrtanik[79, 78], M. Zavrtanik[78, 79], I. Zaw[96], A. Zepeda[64], M. Zimbres-Silva[20, 38] M. Ziolkowski[45]

[1] *Centro Atómico Bariloche and Instituto Balseiro (CNEA- UNCuyo-CONICET), San Carlos de Bariloche, Argentina*
[2] *Centro Atómico Constituyentes (Comisión Nacional de Energía Atómica/CONICET/UTN-FRBA), Buenos Aires, Argentina*
[3] *Centro de Investigaciones en Láseres y Aplicaciones, CITEFA and CONICET, Argentina*
[4] *Departamento de Física, FCEyN, Universidad de Buenos Aires y CONICET, Argentina*
[7] *IFLP, Universidad Nacional de La Plata and CONICET, La Plata, Argentina*
[8] *Instituto de Astronomía y Física del Espacio (CONICET- UBA), Buenos Aires, Argentina*
[9] *Instituto de Física de Rosario (IFIR) - CONICET/U.N.R. and Facultad de Ciencias Bioquímicas y Farmacéuticas U.N.R., Rosario, Argentina*
[10] *National Technological University, Faculty Mendoza (CONICET/CNEA), Mendoza, Argentina*
[11] *Observatorio Pierre Auger, Malargüe, Argentina*
[12] *Observatorio Pierre Auger and Comisión Nacional de Energía Atómica, Malargüe, Argentina*
[13] *University of Adelaide, Adelaide, S.A., Australia*
[16] *Centro Brasileiro de Pesquisas Fisicas, Rio de Janeiro, RJ, Brazil*
[17] *Pontifícia Universidade Católica, Rio de Janeiro, RJ, Brazil*

[18] *Universidade de São Paulo, Instituto de Física, São Carlos, SP, Brazil*
[19] *Universidade de São Paulo, Instituto de Física, São Paulo, SP, Brazil*
[20] *Universidade Estadual de Campinas, IFGW, Campinas, SP, Brazil*
[21] *Universidade Estadual de Feira de Santana, Brazil*
[22] *Universidade Estadual do Sudoeste da Bahia, Vitoria da Conquista, BA, Brazil*
[23] *Universidade Federal da Bahia, Salvador, BA, Brazil*
[24] *Universidade Federal do ABC, Santo André, SP, Brazil*
[25] *Universidade Federal do Rio de Janeiro, Instituto de Física, Rio de Janeiro, RJ, Brazil*
[26] *Universidade Federal Fluminense, EEIMVR, Volta Redonda, RJ, Brazil*
[27] *Rudjer Bošković Institute, 10000 Zagreb, Croatia*
[28] *Charles University, Faculty of Mathematics and Physics, Institute of Particle and Nuclear Physics, Prague, Czech Republic*
[29] *Institute of Physics of the Academy of Sciences of the Czech Republic, Prague, Czech Republic*
[30] *Palacky University, RCATM, Olomouc, Czech Republic*
[32] *Institut de Physique Nucléaire d'Orsay (IPNO), Université Paris 11, CNRS-IN2P3, Orsay, France*
[33] *Laboratoire AstroParticule et Cosmologie (APC), Université Paris 7, CNRS-IN2P3, Paris, France*
[34] *Laboratoire de l'Accélérateur Linéaire (LAL), Université Paris 11, CNRS-IN2P3, Orsay, France*
[35] *Laboratoire de Physique Nucléaire et de Hautes Energies (LPNHE), Universités Paris 6 et Paris 7, CNRS-IN2P3, Paris, France*
[36] *Laboratoire de Physique Subatomique et de Cosmologie (LPSC), Université Joseph Fourier, INPG, CNRS-IN2P3, Grenoble, France*
[37] *SUBATECH, École des Mines de Nantes, CNRS-IN2P3, Université de Nantes, Nantes, France*
[38] *Bergische Universität Wuppertal, Wuppertal, Germany*
[39] *Karlsruhe Institute of Technology - Campus North - Institut für Kernphysik, Karlsruhe, Germany*
[40] *Karlsruhe Institute of Technology - Campus North - Institut für Prozessdatenverarbeitung und Elektronik, Karlsruhe, Germany*
[41] *Karlsruhe Institute of Technology - Campus South - Institut für Experimentelle Kernphysik (IEKP), Karlsruhe, Germany*
[42] *Max-Planck-Institut für Radioastronomie, Bonn, Germany*
[43] *RWTH Aachen University, III. Physikalisches Institut A, Aachen, Germany*
[44] *Universität Hamburg, Hamburg, Germany*
[45] *Universität Siegen, Siegen, Germany*
[46] *Dipartimento di Fisica dell'Università and INFN, Genova, Italy*
[47] *Università dell'Aquila and INFN, L'Aquila, Italy*
[48] *Università di Milano and Sezione INFN, Milan, Italy*
[49] *Dipartimento di Fisica dell'Università del Salento and Sezione INFN, Lecce, Italy*
[50] *Università di Napoli "Federico II" and Sezione INFN, Napoli, Italy*
[51] *Università di Roma II "Tor Vergata" and Sezione INFN, Roma, Italy*
[52] *Università di Catania and Sezione INFN, Catania, Italy*
[53] *Università di Torino and Sezione INFN, Torino, Italy*
[54] *Dipartimento di Ingegneria dell'Innovazione dell'Università del Salento and Sezione INFN, Lecce, Italy*
[56] *Istituto di Astrofisica Spaziale e Fisica Cosmica di Palermo (INAF), Palermo, Italy*
[57] *Istituto di Fisica dello Spazio Interplanetario (INAF), Università di Torino and Sezione INFN, Torino, Italy*
[58] *INFN, Laboratori Nazionali del Gran Sasso, Assergi (L'Aquila), Italy*
[61] *Università di Palermo and Sezione INFN, Catania, Italy*
[63] *Benemérita Universidad Autónoma de Puebla, Puebla, Mexico*
[64] *Centro de Investigación y de Estudios Avanzados del IPN (CINVESTAV), México, D.F., Mexico*
[66] *Universidad Michoacana de San Nicolas de Hidalgo, Morelia, Michoacan, Mexico*
[67] *Universidad Nacional Autonoma de Mexico, Mexico, D.F., Mexico*
[68] *IMAPP, Radboud University Nijmegen, Netherlands*
[69] *Kernfysisch Versneller Instituut, University of Groningen, Groningen, Netherlands*
[70] *Nikhef, Science Park, Amsterdam, Netherlands*
[71] *ASTRON, Dwingeloo, Netherlands*
[72] *Institute of Nuclear Physics PAN, Krakow, Poland*

[73] University of Łódź, Łódź, Poland
[74] LIP and Instituto Superior Técnico, Lisboa, Portugal
[78] J. Stefan Institute, Ljubljana, Slovenia
[79] Laboratory for Astroparticle Physics, University of Nova Gorica, Slovenia
[80] Instituto de Física Corpuscular, CSIC-Universitat de València, Valencia, Spain
[81] Universidad Complutense de Madrid, Madrid, Spain
[82] Universidad de Alcalá, Alcalá de Henares (Madrid), Spain
[83] Universidad de Granada & C.A.F.P.E., Granada, Spain
[84] Universidad de Santiago de Compostela, Spain
[85] Rudolf Peierls Centre for Theoretical Physics, University of Oxford, Oxford, United Kingdom
[87] School of Physics and Astronomy, University of Leeds, United Kingdom
[88] Argonne National Laboratory, Argonne, IL, USA
[89] Case Western Reserve University, Cleveland, OH, USA
[90] Colorado School of Mines, Golden, CO, USA
[91] Colorado State University, Fort Collins, CO, USA
[92] Colorado State University, Pueblo, CO, USA
[93] Fermilab, Batavia, IL, USA
[94] Louisiana State University, Baton Rouge, LA, USA
[95] Michigan Technological University, Houghton, MI, USA
[96] New York University, New York, NY, USA
[97] Northeastern University, Boston, MA, USA
[98] Ohio State University, Columbus, OH, USA
[99] Pennsylvania State University, University Park, PA, USA
[100] Southern University, Baton Rouge, LA, USA
[101] University of Chicago, Enrico Fermi Institute, Chicago, IL, USA
[105] University of Nebraska, Lincoln, NE, USA
[106] University of New Mexico, Albuquerque, NM, USA
[108] University of Wisconsin, Madison, WI, USA
[109] University of Wisconsin, Milwaukee, WI, USA
[110] Institute for Nuclear Science and Technology (INST), Hanoi, Vietnam
[†] Deceased
[a] at Konan University, Kobe, Japan

# Update on the measurement of the CR energy spectrum above $10^{18}$ eV made using the Pierre Auger Observatory

FRANCESCO SALAMIDA[1] FOR THE PIERRE AUGER COLLABORATION[2]
[1]*Università dell'Aquila and INFN, L'Aquila, Italy*
[2]*Observatorio Pierre Auger, Av. San Martin Norte 304, 5613 Malargüe, Argentina*
*(Full author list: http://www.auger.org/archive/authors_2011_05.html)*
*auger_spokespersons@fnal.gov*

**Abstract:** The flux of cosmic rays above $10^{18}$ eV has been measured with unprecedented precision at the Pierre Auger Observatory. Two analysis techniques have been used to extend the spectrum downwards from $3 \times 10^{18}$ eV, with the lower energies being explored using a unique technique that exploits the hybrid strengths of the instrument. The spectral features are presented in detail and the impact of systematic uncertainties on these features is addressed. The increased exposure of about 60% with respect to previous publications is exploited.

## 1 Introduction

In this paper we present an updated measurement of the cosmic ray energy spectrum with the Pierre Auger Observatory [1]. An accurate measurement of the cosmic ray flux above $10^{18}$ eV is a crucial aid for discriminating between different models describing the transition between galactic and extragalactic cosmic rays, the suppression induced by the cosmic ray propagation and the features of the injection spectrum at the sources.

Two complementary techniques are used at the Pierre Auger Observatory to study extensive air showers initiated by ultra-high energy cosmic rays (UHECR): a *surface detector array* (SD) and a *fluorescence detector* (FD).

The SD consists of an array of over 1600 water Cherenkov detectors covering an area of about 3000 km$^2$ allowing the sampling of electrons, photons and muons in the air showers at ground level with an on-time of almost 100%. In addition the atmosphere above the surface detector is observed during clear, moonless nights by 27 optical telescopes grouped in 5 buildings. This detector is used to observe the longitudinal development of an extensive air shower by detecting the fluorescence light emitted by excited nitrogen molecules and the Cherenkov light induced by the particles in the shower. Details regarding the design and the status of the Observatory can be found elsewhere [2, 3, 4].

The energy spectrum at energies greater than $3 \times 10^{18}$ eV has been derived using data from the surface detector array of the Pierre Auger Observatory. The analysis of air showers measured with the fluorescence detector that triggered at least one station of the surface detector array (i.e. *hybrid* events) enables measurements to be extended to lower energies. Despite the limited number of events due to the fluorescence detector on-time, the lower energy threshold and the good energy resolution of *hybrid* events allow us to measure the flux of cosmic rays down to $10^{18}$ eV in the energy region where the transition between galactic and extragalactic cosmic rays is expected [5, 6, 7].

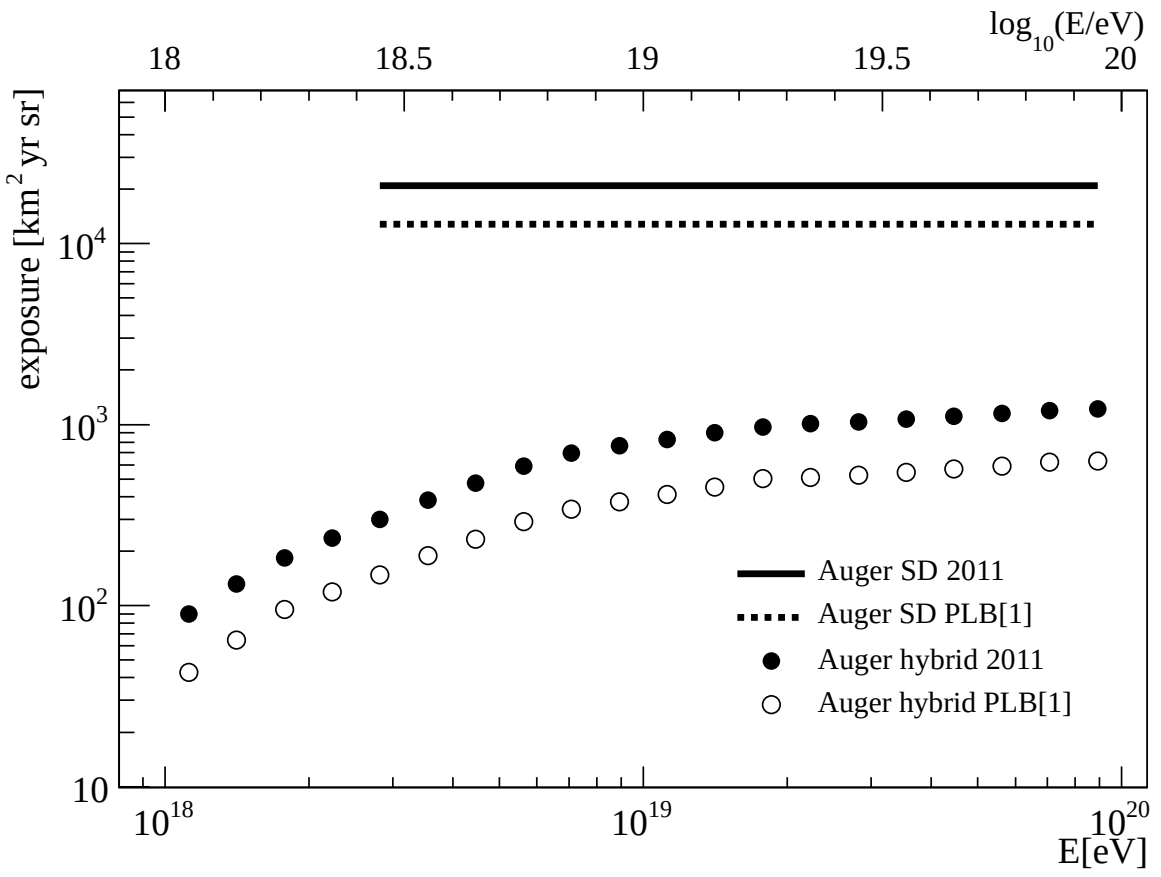

Figure 1: The SD and hybrid exposures used for the current flux measurement compared with a previously published data set [1]. The SD exposure is shown for energies higher than $10^{18.5}$ eV where the detector is fully efficient.

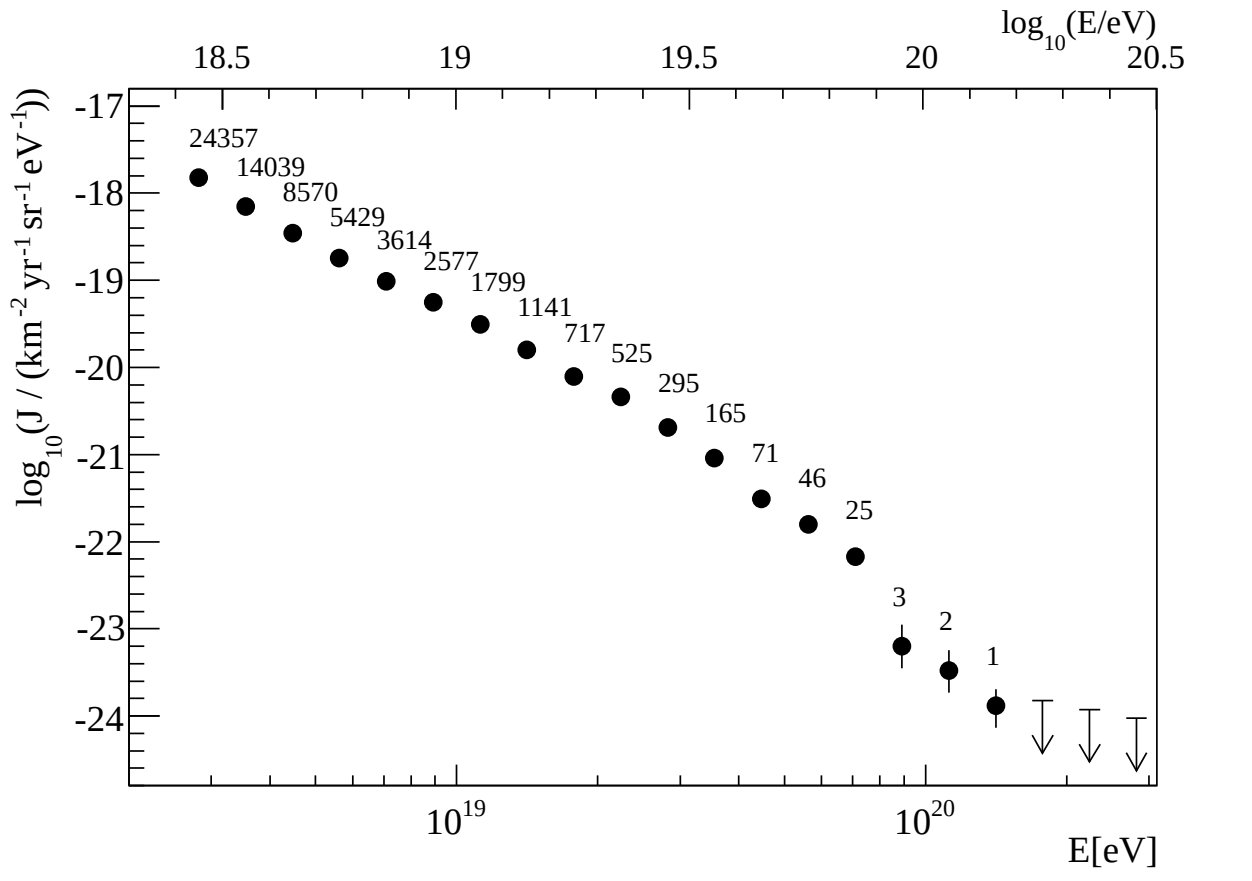

Figure 2: Energy spectrum derived from surface detector data calibrated with fluorescence detector measurements. The spectrum has been corrected for the energy resolution of the detector. Only statistical uncertainties are shown. Upper limits correspond to 68% CL.

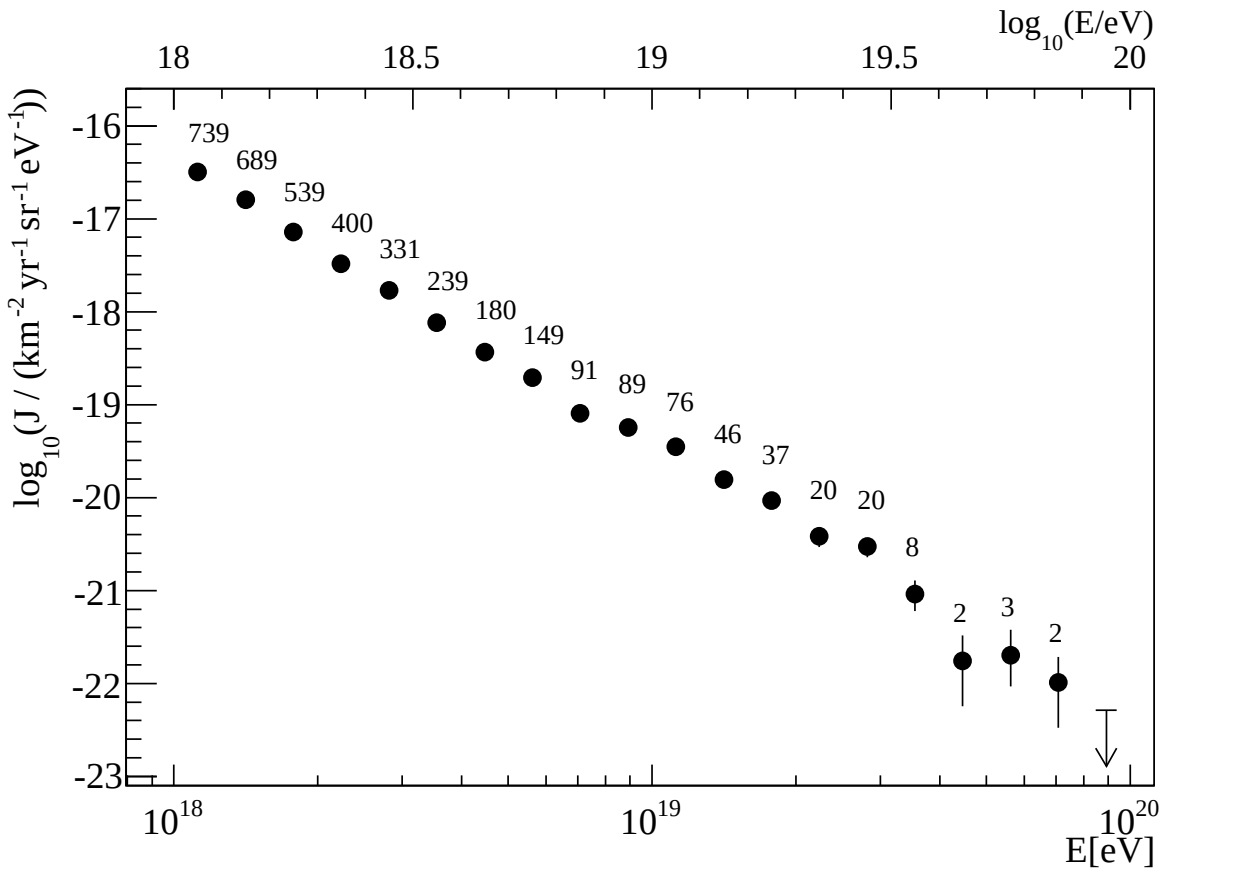

Figure 3: Energy spectrum derived from hybrid data. Only statistical uncertainties are shown. Upper limits correspond to 68% CL.

## 2 Surface detector spectrum

Here we report an update of the energy spectrum based on the surface detector data [1] using the period between 1 January 2004 and 31 December 2010. The exposure increased by about 60% with respect to the previous publication and is now 20905 km$^2$ sr yr. It is calculated by integrating the number of active detector stations of the surface array over time. The SD exposure is shown in Fig. 1 compared to the one used in [1]. Above $3 \times 10^{18}$ eV the SD acceptance is saturated regardless of the primary mass. The uncertainty on the derivation of the exposure is about 3% [8].

The event selection requires the water-Cherenkov detector with the greatest signal to be surrounded by operational stations and the reconstructed zenith angle to be smaller than 60°. The total number of events above $3 \times 10^{18}$ eV fulfilling the selection criteria is about 64000. The number of events with energy greater than $10^{19}$ eV is about 5000. The number of events above $3 \times 10^{18}$ eV does not fully reflect the increase in exposure with respect to previous publication as the energy calibration has changed meantime [9].

As the energy estimator for the SD we use the expected signal at 1000 m from the shower core, corrected for shower attenuation effects. The calibration of the energy estimator of the surface detector is based on events measured in coincidence with the fluorescence detector [9]. The procedure is affected by a systematic error of 22% due to the uncertainty on the fluorescence energy assignment.

The energy resolution of the SD is $\sim$16% at threshold, falling to $\sim$12% above 10 EeV. Details can be found in [9]. The influence of the bin-to-bin migration on the reconstruction of the flux due to the energy resolution has been corrected by applying a forward-folding approach. The correction of the flux is mildly energy dependent but is less than 20% over the entire energy range.

The energy spectrum, including the correction of the energy resolution, is shown in Fig. 2. The number of events of the raw distribution is superimposed. The total systematic uncertainty of the flux for the derived spectrum is 6% and is obtained by summing in quadrature the exposure uncertainty (3%) and that due to the forward-folding assumptions (5%).

## 3 Hybrid energy spectrum

The energy spectrum from hybrid events is determined from data taken between 1 November 2005 and 30 September 2010. With respect to the previous publication [1] the time period has been extended and the events recorded at the site of the Loma Amarilla fluorescence building, the final set of telescopes brought into operation, have been added into the analysis. The resulting integrated exposure is doubled with respect to the previous publication [1, 10]. To ensure good energy reconstruction only events that satisfy strict quality criteria have been accepted [10]. In particular, to avoid a possible bias in event selection due to the differences between shower profiles initiated by primaries of different mass, only showers with geometries that would allow the observation of all primaries in the range from proton to iron are retained in the data sample. The corresponding fiducial volume in terms of shower-telescope distance and zenith angle range is defined as a function of the reconstructed energy and has been verified with data [11]. A detailed simulation of the detector response has shown that for zenith angles less than 60°, every FD trigger above $10^{18}$ eV passing all the selection criteria is accompanied by a SD trigger of at least one station, independent of the mass or direction of the incoming primary particle [10].

The exposure of the hybrid mode of the Pierre Auger Observatory has been calculated using a time-dependent

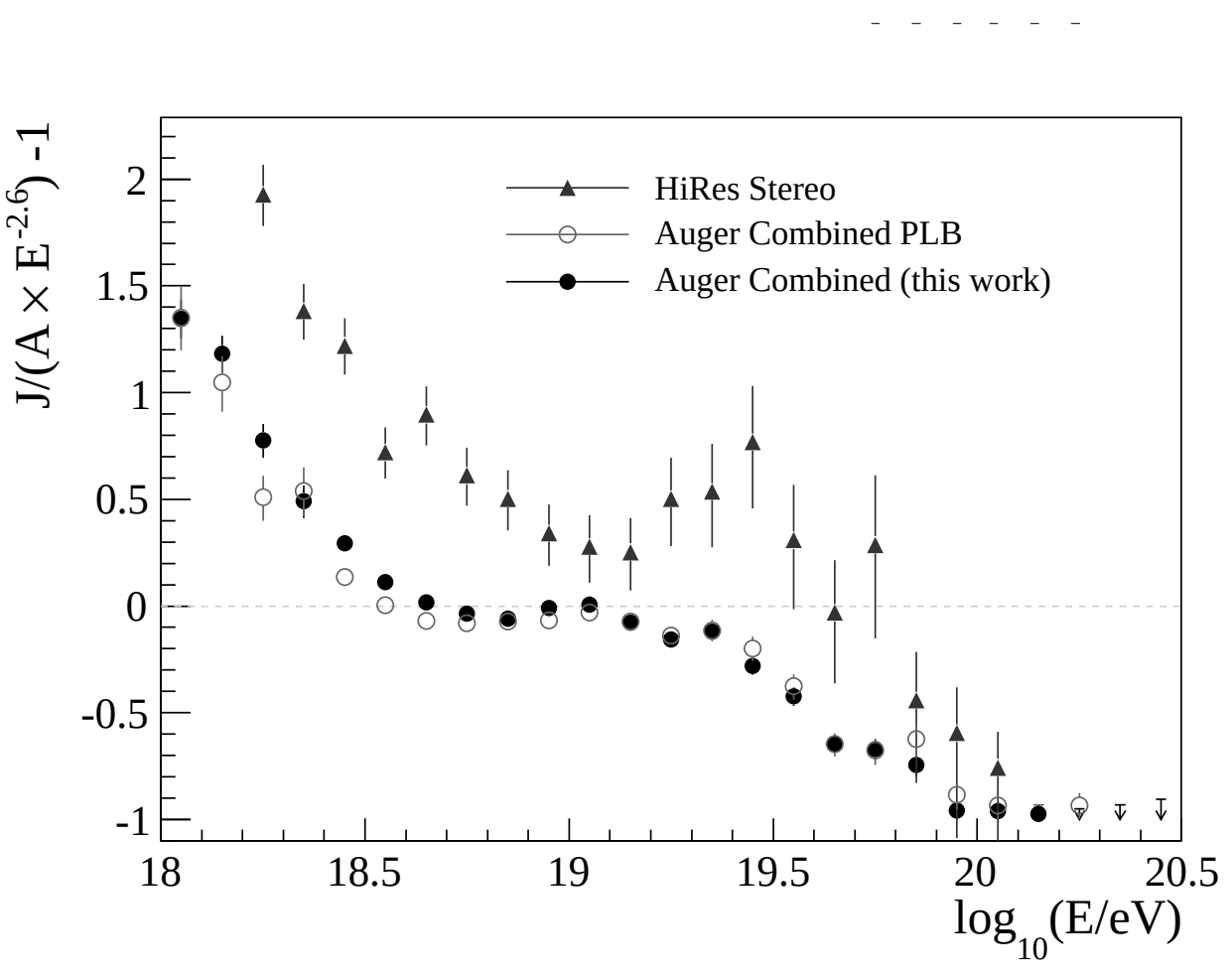

Figure 4: Fractional difference between the combined energy spectrum of the Pierre Auger Observatory and a spectrum with an index of 2.6. Data from HiRes stereo measurements [20] are shown for comparison.

Monte Carlo simulation. The changing configurations of both fluorescence and surface detectors are taken into account for the determination of the on-time of the hybrid system. Within a time interval of 10 min, the status and efficiency of all detector components of the Observatory, down to the level of the single PMTs of the fluorescence detector, are determined. Moreover, all atmospheric measurements [12] as well as monitoring information are considered and used as input for the simulation. A detailed description can be found in [10, 13]. The longitudinal profiles of the energy deposits have been simulated with the CONEX [14] air shower simulation program with Sibyll 2.1 [15] and QGSJet II-0.3 [16] as alternative hadronic interaction models. The influence of the assumptions made in the hadronic interaction models on the exposure calculation has been estimated to be lower than 2%. A 50% mixture of protons and iron nuclei has been assumed for the primaries. The quality cuts used for the event selection lead to only a small dependence of the exposure on the mass composition. The systematic uncertainty arising from the lack of knowledge of the mass composition is about 8% (1%) at $10^{18}$ eV ($> 10^{19}$ eV). The full MC simulation chain has been cross-checked with air shower observations and the analysis of laser shots fired from the Central Laser Facility [17]. The total systematic uncertainty of the derived exposure is estimated as 10% (6%) at $10^{18}$ eV ($> 10^{19}$ eV).

The energy spectrum calculated using the hybrid events is shown in Fig. 3. The main systematic uncertainty is due to the energy assignment which relies on the knowledge of the fluorescence yield, choice of models and mass composition [18], absolute detector calibration [19] and shower reconstruction. The total uncertainty is estimated to be about 22%. The details can be found in [1].

Table 1: Fitted parameters and their statistical uncertainties characterizing the combined energy spectrum.

| parameter | broken power laws | power laws + smooth function |
|---|---|---|
| $\gamma_1(E < E_{\rm ankle})$ | $3.27 \pm 0.02$ | $3.27 \pm 0.01$ |
| $\lg(E_{\rm ankle}/{\rm eV})$ | $18.61 \pm 0.01$ | $18.62 \pm 0.01$ |
| $\gamma_2(E > E_{\rm ankle})$ | $2.68 \pm 0.01$ | $2.63 \pm 0.02$ |
| $\lg(E_{\rm break}/{\rm eV})$ | $19.41 \pm 0.02$ | |
| $\gamma_3(E > E_{\rm break})$ | $4.2 \pm 0.1$ | |
| $\lg(E_{1/2}/{\rm eV})$ | | $19.63 \pm 0.02$ |
| $\lg(W_{\rm c}/{\rm eV})$ | | $0.15 \pm 0.02$ |
| $\chi^2/{\rm ndof}$ | $37.8/16 = 2.7$ | $33.7/16 = 2.3$ |

## 4 Combined energy spectrum

The energy spectrum derived from hybrid data has been combined with the one obtained from surface detector data using a maximum likelihood method. Since the surface detector energy estimator is calibrated with hybrid events [9], the two spectra have the same systematic uncertainty in the energy scale (22%). On the other hand, the normalisation uncertainties are independent. They are taken as 6% for the SD and 10% (6%) for the hybrid flux at $10^{18}$ eV ($> 10^{19}$ eV). These normalisation uncertainties are used as additional constraints in the combination. This combination procedure is used to derive the scale parameters $k_{\rm SD}$=1.01 and $k_{\rm FD}$=0.99 which have to be applied to the individual spectra in order to match them. The fractional difference of the combined energy spectrum with respect to an assumed flux $\propto E^{-2.6}$ is shown in Fig. 4. The measurements in stereo mode from the HiRes experiment [20] are also shown in Fig. 4 for comparison. The ankle feature seems to be somewhat more sharply defined in the Auger data. This is possibly due to the different energy resolution of the two instruments. A comparison with the Auger flux published in [1] is also shown in Fig. 4. The two spectra are compatible within the systematic uncertainties. Furthermore, it has to be noted that the updated spectrum includes the change in the calibration curve reported in [9].

The characteristic features of the combined spectrum have been quantified in two ways. For the first method, shown as a dotted line in Fig. 5, three power laws with free breaks between them have been used. For the second approach, two power laws in the ankle region and a smoothly changing function at higher energies have been adopted. The function is given by

$$J(E; E > E_{\rm ankle}) \quad \propto \quad E^{-\gamma_2} \, \frac{1}{1 + \exp\left(\frac{\lg E - \lg E_{1/2}}{\lg W_c}\right)},$$

where $E_{1/2}$ is the energy at which the flux has fallen to one half of the value of the power-law extrapolation and $W_c$ parameterizes the width of the transition region. The result of the fit is shown as black solid line in Fig. 5. The derived parameters quoting only the statistical uncertainties are given in Table 1. Changes to the calibration curve [9] have re-sulted in some changes of the parameters of the spectrum with respect to previous work [1], although only the values of $\gamma_2$ are different by more than the quoted statistical uncertainties (in Ref. [1] a value of $\gamma_2 = 2.59 \pm 0.02$ is reported).

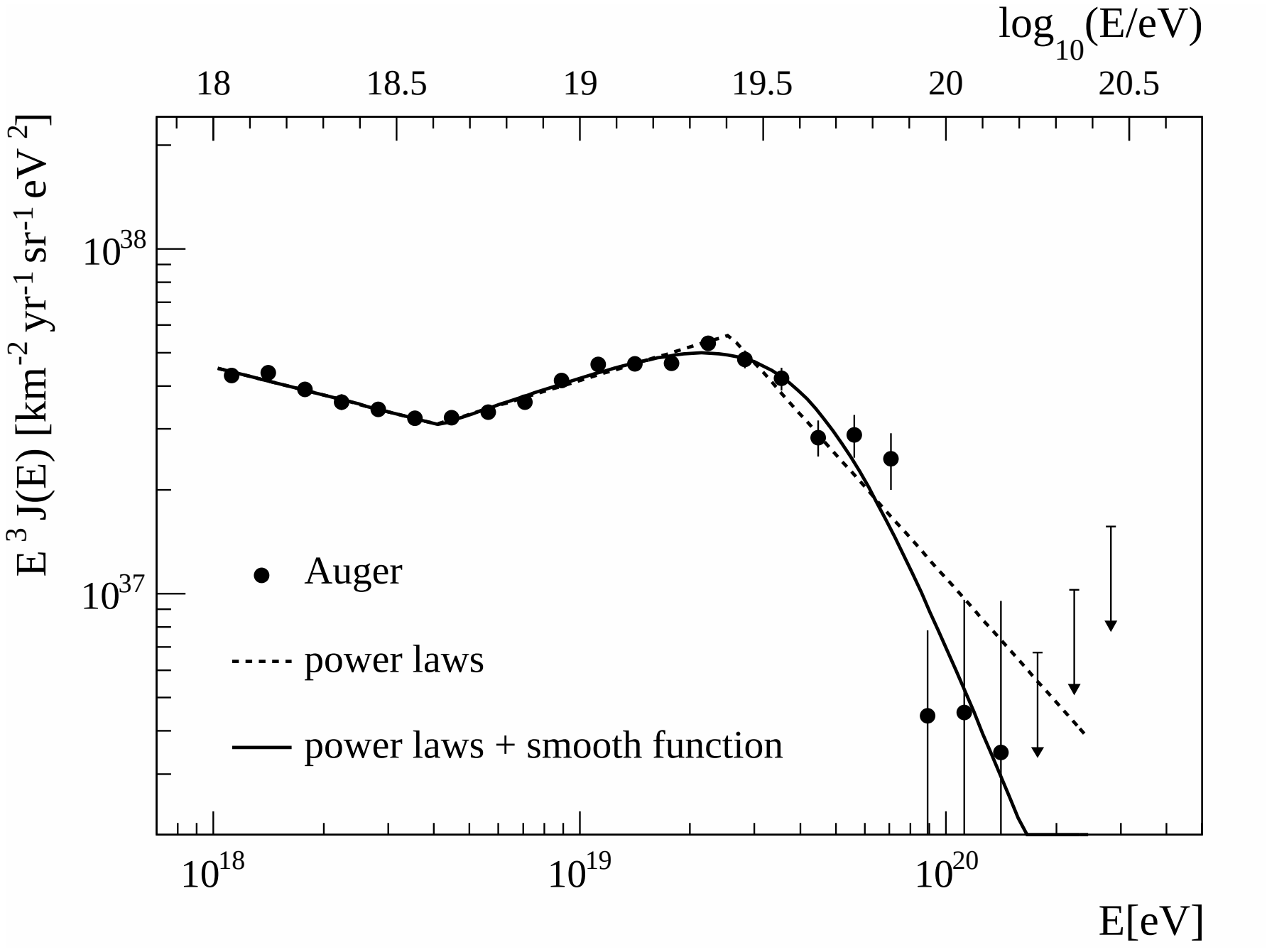

Figure 5: The combined energy spectrum is fitted with two functions (see text). Only statistical uncertainties are shown. The systematic uncertainty in the energy scale is 22%.

## 5 Summary

An update of the measurement of the cosmic ray flux with the Pierre Auger Observatory has been presented. Two independent measurements of the cosmic ray energy spectrum with the Pierre Auger Observatory have been exploited. Both spectra share the same systematic uncertainties in the energy scale. A combined spectrum has been derived with high statistics covering the energy range from $10^{18}$ eV to above $10^{20}$ eV. The dominant systematic uncertainty of the spectrum stems from that of the overall energy scale, which is estimated to be 22%. The combination of spectra enables the precise measurement of both the ankle and the flux suppression at highest energies.

# The Cosmic Ray Spectrum above $4 \times 10^{18}$ eV as measured with inclined showers recorded at the Pierre Auger Observatory

HANS P. DEMBINSKI[1] FOR THE PIERRE AUGER COLLABORATION[2]
[1]*Karlsruhe Institute of Technology (KIT), IEKP, Karlsruhe, Germany*
[2]*Observatorio Pierre Auger, Av. San Martin Norte 304, 5613 Malargüe, Argentina*
*(Full author list: http://www.auger.org/archive/authors_2011_05.html)*
*auger_spokespersons@fnal.gov*

**Abstract:** The energy spectrum above $4\times10^{18}$ eV is presented, obtained from 5936 events with zenith angles exceeding $62°$ collected by the Surface Detector of the Pierre Auger Observatory from 1 January 2004 to 31 December 2010. Showers with such large zenith angles are muon-dominated at ground level and the radial symmetry around the shower axis is broken due to geomagnetic deflections. They are analysed separately from showers with smaller zenith angles using two-dimensional models of the muon density at ground, allowing one to reconstruct a global muon number for every event. The conversion of the muon number to energy is obtained using the sub-sample of events detected simultaneously with both the Surface and the Fluorescence Detector. The spectrum obtained displays suppression near $4 \times 10^{19}$ eV compatible with the analysis that uses less inclined events.

## 1 Introduction

The Pierre Auger Observatory uses two techniques to study cosmic rays, exploiting the induced extensive air showers in Earth's atmosphere. Charged particles and photons which arrive at ground are measured with more than 1600 water-Cherenkov detectors, most of which are on a 1.5 km triangular-grid distributed over 3000 km$^2$ (Surface Detector Array, SD [1]). In addition, charged particles in the air generate ultra-violet light by excitation of nitrogen, which is observed by 27 fluorescence telescopes (Fluorescence Detector, FD [2]) under suitable conditions.

The SD has a duty cycle of almost 100 % and collects the main bulk of events. Its energy scale is derived from coincident measurements with the FD, which provides an almost calorimetric energy estimate of the shower [3]. The FD can only operate in dark, moonless nights with a field of view free of clouds. This limits its duty cycle to 13 % [4].

About $1/4$ of the collected air showers have zenith angles exceeding $60°$. These *very inclined showers* are reconstructed separately from less inclined ones due to their special phenomenology. Very inclined showers are muon-dominated at ground and show a broken circular symmetry in the lateral fall-off of particle density, partly due to deflections in the geomagnetic field and partly due to the different trajectories of early and late arriving particles. Only a weak halo of low energy electrons and photons, generated mainly by muon decay, arrives with the muons. Its contribution to the SD signals is typically small and well understood [5].

Very inclined showers are interesting, because they increase the viewable portion of the sky and the event statistics. Moreover, they allow one to study the muon component of air showers under weak model assumptions [6].

In this proceeding, we give an update of the cosmic ray flux obtained from inclined events [7, 8] collected from 1 January 2004 to 31 December 2010. The analysis is based on 5936 events above $4 \times 10^{18}$ eV, the lowest energy where the SD is fully efficient in the zenith angle range $62° < \theta < 80°$. Our main improvements are an extensive validation of the reconstruction chain with air shower simulations [6], leading to reliable estimates of the reconstruction uncertainties, and the switch to a maximum likelihood method for the energy calibration.

## 2 Event selection and reconstruction

Very inclined showers generate sparse and elongated signal patterns on the SD with a sharp rise of the signal in time, typical for a front of arriving muons [9]. Surface detectors trigger on such signals. The central data acquisition builds events from these local triggers if they have a compact spatial pattern and arrival times that roughly agree with a plane moving with the speed of light across the array [10].

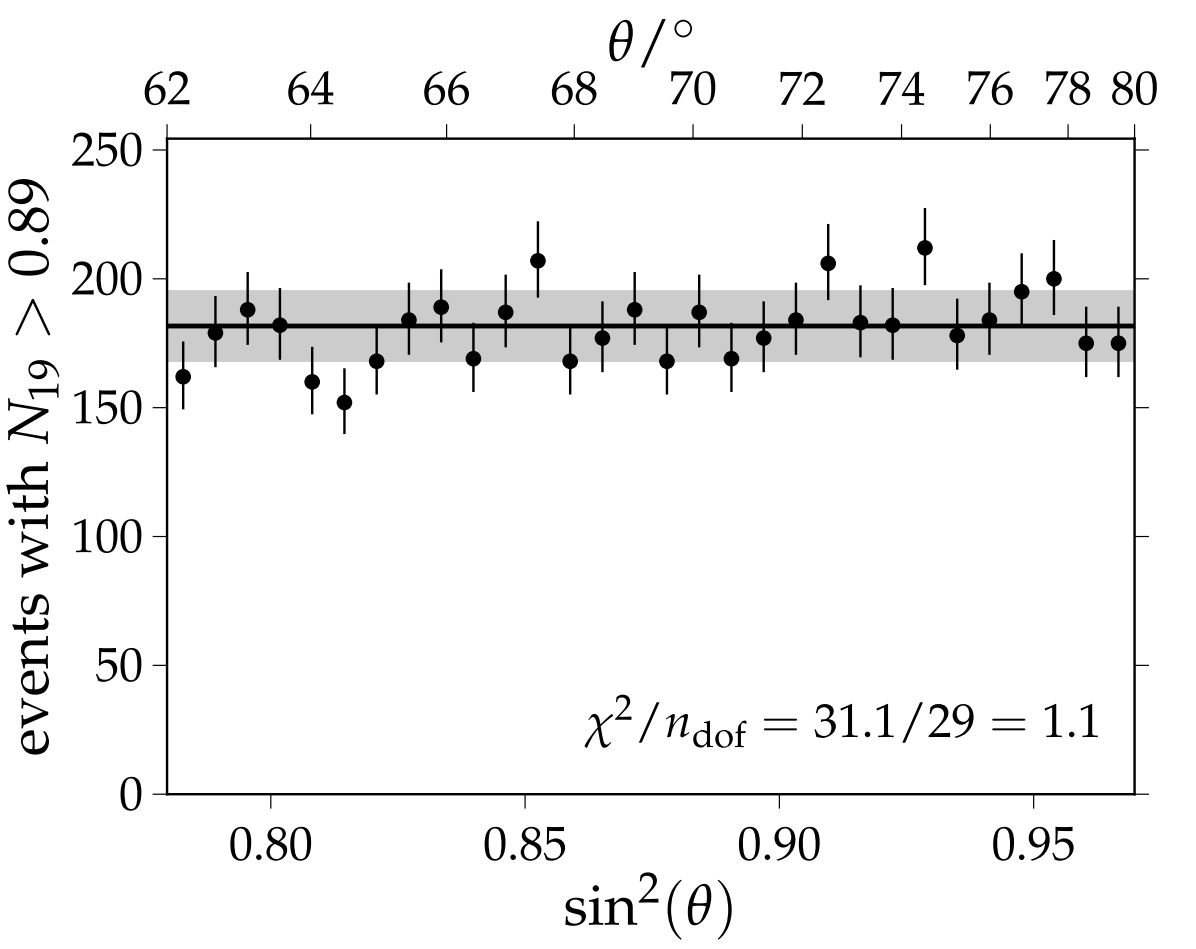

Figure 1: Number of recorded events with $N_{19} > 0.89$ (corresponding to the threshold of full SD efficiency) in bins of equal geometrical exposure. A fitted horizontal line is compared and the reduced $\chi^2$ is shown.

Muons from showers of lower energy form a background which generates false triggers or false early signals in individual detectors. This background is removed off-line by requiring a stricter space-time compatibility of the event with a plane front, using an algorithm that successively tries all combinations of rejecting one or more stations until an acceptable configuration is found. False early signals can be identified as small isolated and narrow peaks before the main signal cluster in the time domain and are rejected with a heuristic algorithm based on that property.

The arrival direction $(\theta, \phi)$ of the cosmic ray is reconstructed from the signal arrival times by fitting a sphere expanding with the speed of light from a point along the shower axis. The achieved angular resolution is better than 1° above $4 \times 10^{18}$ eV [6, 9, 11]. The average radius of curvature for a shower of zenith angle 70° is around 35 km.

In the next step, the shower size on the ground is reconstructed: it scales with energy $E$ of the cosmic ray. For that purpose, the measured signals are compared to expected signals, which are computed as follows. Firstly, the number of muons hitting a surface detector is calculated from a model of the lateral fall-off of muon density as a function of the incoming direction of the shower. The model accounts for circular asymmetries and the muon attenuation with the zenith angle. Such models have been derived from simulations using different approaches [12, 13], with comparable results. Then, the detector response to incoming muons is calculated, based on GEANT4 simulations. Finally, the expected signal from electromagnetic particles is added, parametrized from simulations [5]. At zenith angles larger than 60° these particles are mainly generated by muon decay and contribute about 15 % to the signal.

The shower core and shower size are simultaneously estimated with a maximum likelihood method that accounts for non–triggering and saturated detectors. The shower size parameter $N_{19}$ is proportional to the total number of muons in the shower and scales only with the cosmic ray energy and mass. The model of the lateral fall-off of muon density is normalised in such a way that $N_{19}$ = 1 indicates a shower with the same number of muons as a simulated shower initiated by a proton of $10^{19}$ eV. However, $N_{19}$ is not an absolute quantity, it depends on the hadronic interaction model used to simulate the proton shower. The model used here [13] is based on QGSJet-II [14] and yields values of $N_{19}$ that are about 10 % higher than in previous analyses [8]. This change does not affect the energy determination, as it is reabsorbed in the calibration procedure (see section 3).

The reconstruction chain was validated with an analysis of more than 100,000 simulated SD events which allows one to assess bias and resolution of the reconstruction. Further details are given in [6]. Above $4 \times 10^{18}$ eV, the resolution of $N_{19}$ is better than 20 % and the systematic uncertainty smaller than 3 %.

Finally, a fiducial cut on the SD area is defined to guarantee a high-quality reconstruction of the events in the zenith angle range considered for the spectrum. Events only pass if the station nearest to the reconstructed core has six active neighbours in the surrounding hexagon. Above the energy where the SD is fully efficient in the considered zenith angle range, the exposure is the integral over unit cells of detectors that pass the fiducial cut, time, and solid angle. We checked the distribution of $\sin^2\theta$ of events for different cuts in $N_{19}$ and zenith angle. For $62° < \theta < 80°$ and $N_{19} > 0.89$ corresponding to $4 \times 10^{18}$ eV, the distribution becomes flat as expected in the regime of full efficiency. This is shown in figure 1. Under these conditions, the integrated exposure over the period considered amounts to 5306 km$^2$ sr yr, with a systematic uncertainty of 3 % [10]. Disregarding the range $60° < \theta < 62°$ allowed us to reduce the energy threshold by 40 % with respect to our previous report [8].

## 3 Energy calibration

A high-quality selection of events observed simultaneously with FD and SD is used to calibrate the shower size parameter $N_{19}$. In addition to the cuts on the SD described already, we require $\sigma[N_{19}]/N_{19} < 0.2$. For the FD, we look for a good reconstruction of the longitudinal profile: at least 6 triggered pixels, track length $> 200\,\mathrm{g\,cm^{-2}}$, radial distance of the SD detector used in angular reconstruction to the shower axis $< 750$ m, fraction of Cherenkov light $< 50\,\%$, $\chi^2_{\rm GH}/n_{\rm dof} < 2.5$ for a fitted Gaisser-Hillas curve and $(\chi^2_{\rm line} - \chi^2_{\rm GH})/n_{\rm dof} > 4$ for a fitted line, depth of shower maximum $X_{\rm max}$ farther away than $50\,\mathrm{g\,cm^{-2}}$ from the borders of the field of view, $\sigma[X_{\rm max}] < 50\,\mathrm{g\,cm^{-2}}$, and $\sigma[E]/E < 0.2$. Above $4 \times 10^{18}$ eV, 125 events are selected. The cuts are more restrictive than the ones used previously [8] where 145 events were kept even though the data set was smaller, but less restrictive than those in [3].

A power law $N_{19} = A(E/10^{19}\,\mathrm{eV})^B$ is fitted to these events. Its inverse serves as the calibration function. A maximum likelihood method was developed to perform the fit [15] which uses a model of the observed $(E_{\text{FD}}, N_{19}, \theta)$-distribution in form of the probability density function (p.d.f.) $f_{\text{hyb}}$. It is constructed from the following train of thought. The ideal points described by the power law are not uniformly distributed in energy, but follow a steeply falling distribution $h(E, \theta)$ given by the cosmic ray flux multiplied with the FD fiducial area. Shower-to-shower fluctuations described by a Gaussian p.d.f. $g_{\text{sh}}$ shift $N_{19}$ away from the curve. Further shifts are added by sampling fluctuations of the SD and FD described by the Gaussian p.d.f.s $g_{\text{FD}}$ and $g_{\text{SD}}$. Eventually, a point may disappear due to the limited SD efficiency $\epsilon_{\text{SD}}$. Mathematically, the whole process is described by the convolution integral

$$f_{\text{hyb}}(E_{\text{FD}}, N_{19}, \theta) = C \times \int \mathrm{d}E \int \mathrm{d}\tilde{N}_{19}\, h(E, \theta) \quad (1)$$
$$g_{\text{FD}}(E_{\text{FD}}|E)\, \epsilon_{\text{SD}}(N_{19}, \theta)\, g_{\text{SD}}(N_{19}|\tilde{N}_{19})\, g_{\text{sh}}\big(\tilde{N}_{19}|\bar{N}_{19}(E)\big),$$

where $\bar{N}_{19}(E)$ is the average value of $N_{19}$ predicted by the power law, $\tilde{N}_{19}$ the shower-to-shower fluctuated version, $N_{19}$ the observed value, and $C$ a normalization constant. The convolution is carried out numerically. Parameters of $h(E, \theta)$, $g_{\text{FD}}$, and $g_{\text{SD}}$ are fitted separately [15]. The FD fluctuations have a constant width of about 8 % above $10^{18}$ eV. The width of the SD fluctuations is described by the function $\sigma[N_{19}]/N_{19} = p_0 + p_1 N_{19}{}^{-1/2}$ with constants $p_0, p_1$. For the energies considered here, $\epsilon_{\text{SD}}$ = 1. The main advantage over the least-squares method, used for example in [8], is the possibility of including data where $\epsilon_{\text{SD}} < 1$, although we do not use this feature.

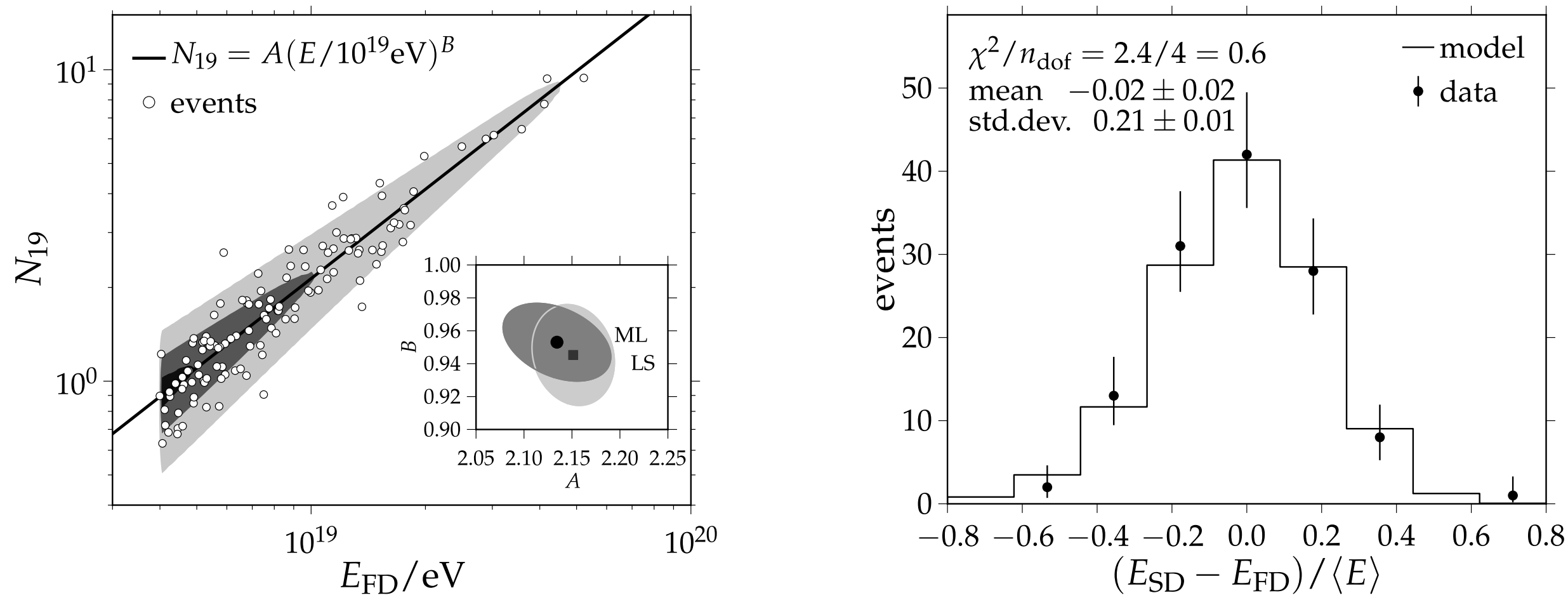

Figure 2: Left: Fit of the calibration curve $N_{19} = A(E/10^{19}\text{eV})^B$ to 125 events. The contours indicate constant levels of the p.d.f. $f_{\text{hyb}}$ (see text) integrated over zenith angle, corresponding to $10, 50, 90$ % of the maximum value. The calibration constants $A, B$ obtained with the maximum-likelihood method (ML) and the former least-squares method (LS) are shown in the inset. The ellipses indicate uncertainty contours of 68 % confidence. Right: Distribution of the difference between calibrated SD energy $E_{\text{SD}}$ and FD energy $E_{\text{FD}}$ divided by their average $\langle E \rangle$. The distribution expected from the model $f_{\text{hyb}}$ is compared. The reduced $\chi^2$ value, mean, and standard deviation of the distribution are given.

The fit of the calibration curve is depicted in figure 2 (left panel). The fitted constants are $A = (2.13 \pm 0.04 \pm 0.11\,(\text{sys.}))$ and $B = (0.95 \pm 0.02 \pm 0.03\,(\text{sys.}))$. The systematic uncertainties are estimated from simulation studies and variations of the FD cuts. The distribution of the relative differences between calibrated SD energies and FD energies (shown in fig. 2 right) is unbiased and agrees with the prediction obtained from $f_{\text{hyb}}$, indicating that $f_{\text{hyb}}$ also describes the fluctuations of 21 % well. The latter are the combination of FD and SD fluctuations and the fitted shower-to-shower fluctuations $\sigma_{\text{sh}}[N_{19}]/N_{19}$. We find $\sigma_{\text{sh}}[N_{19}]/N_{19} = (16 \pm 2)$ % at $4 \times 10^{18}$ eV. The results of the maximum likelihood method agree with those of the former least-squares method (see inset of fig. 2 left), and with previous results [8]. Finally, it is shown in [6] that the calibration curve is discordant with the predictions from Monte-Carlo calculations in the sense that fewer muons are predicted than are observed.

In the zenith angle range $62° < \theta < 65°$, the reconstruction chain used for vertical showers [3, 17] is still applicable and can be used for a crosscheck. The 848 SD events in that zenith angle range above $4 \times 10^{18}$ eV show an average difference of $(2.2 \pm 0.3)$ %, which is within the expected systematic uncertainty of $5\%$.

## 4 Results and discussion

Inclined events recorded from 1 January 2004 to 31 December 2010 were analysed with the procedure outlined above. We obtain 5936 calibrated events in the zenith angle range $62° < \theta < 80°$ above $4 \times 10^{18}$ eV, the energy where the SD becomes fully efficient. Due to the lowered threshold, the number of events is a factor of three larger compared to our previous report [8]. The cosmic ray flux, shown in figure 3, is obtained by dividing the energy spectrum of the cosmic rays by the accumulated exposure of 5306 km$^2$ sr yr in this zenith angle interval.

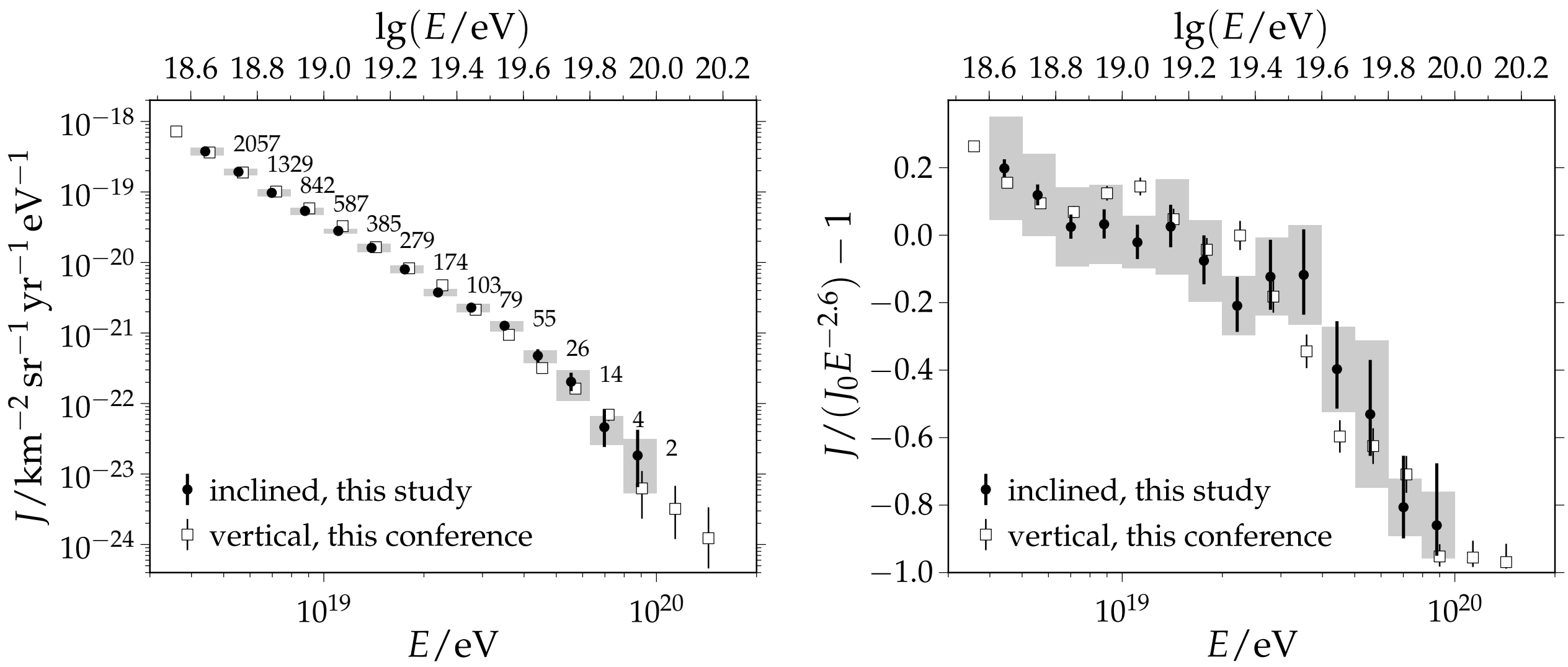

Figure 3: Left: The cosmic ray flux $J(E)$ derived from inclined events with zenith angles $62^{\circ} < \theta < 80^{\circ}$. The numbers indicate the events in each bin. Compared is the flux obtained from vertical events (see [17]). The points have been shifted by a small amount to improve visibility. Right: The shape of $J(E)$ is emphasized by dividing through a reference flux $J_0 E^{-2.6}$ with $J_0 = 9.66 \times 10^{29}\,\mathrm{km}^{-2}\,\mathrm{sr}^{-1}\,\mathrm{yr}^{-1}\,\mathrm{eV}^{1.6}$. Shaded boxes indicate the systematic uncertainty from exposure and energy calibration. The overall energy scale has a systematic uncertainty of 22 % [16].

The uncertainty of the energy calibration was propagated into the flux with a Monte-Carlo approach. The spectrum was re-generated 200 times with calibration constants fluctuated within their uncertainties. The systematic uncertainty of the flux due to exposure and energy calibration is around 13 % up to $4 \times 10^{19}$ eV, and increases up to 70 % at higher energies. The uncertainty of the FD energy scale of 22 % [16] is the largest systematic uncertainty. The flux estimate is slightly distorted by the limited detector resolution. This effect was neglected in this study: when added it will lower the flux estimate slightly.

A power law $E^{-\gamma}$ fitted to the spectrum between $6 \times 10^{18}$ eV and $4 \times 10^{19}$ eV yields a spectral index $\gamma_1 = (2.72 \pm 0.04 \pm 0.04\,(\mathrm{sys.}))$. A flux suppression above $4 \times 10^{19}$ eV is observed, with a sharp break in the spectrum and a new spectral index $\gamma_2 = (4.5 \pm 0.8 \pm 0.04\,(\mathrm{sys.}))$. Both spectral indices agree with the previous values [8] $\gamma_1 = (2.76 \pm 0.06)$ and $\gamma_2 = (5.1 \pm 0.9)$ respectively, and with the values derived from vertical showers [17].

The flux obtained from inclined showers agrees with the one obtained from vertical showers [17] within the systematic uncertainties. If statistical and systematic uncertainties are added in quadrature, the reduced $\chi^2$ value of the flux difference is 10.9/14 = 0.8. The data sets will be combined after further work to reduce the systematic uncertainties and to include the unfolding of detector effects.

# The AMIGA infill detector of the Pierre Auger Observatory: performance and first data

IOANA C. MARIŞ[1] FOR THE PIERRE AUGER COLLABORATION[2]
[1]*Laboratoire de Physique Nucléaire et des Hautes Energies, Universités Paris 6 et Paris 7, CNRS-IN2P3, Paris, France*
[2]*Observatorio Pierre Auger, Av. San Martín Norte 304, 5613 Malargüe, Argentina*
*(Full author list: http://www.auger.org/archive/authors_2011_05.html)*
*auger_spokepersons@fnal.gov*

**Abstract:** We present a first analysis of cosmic rays observed with the AMIGA infill array of the Pierre Auger Observatory. The spacing of 750 m between the surface detectors, half the distance of the regular array, allows us to extend the energy range of interest to energies as low as $3 \times 10^{17}$ eV. The lateral distribution function is presented and the uncertainty of the signal at an optimum distance of 450 m, used to obtain an energy estimator, is discussed. The first steps towards the measurement of the energy spectrum are described. The calculation of the array exposure and the strategy for the energy calibration of the infill, obtained from events observed in coincidence with the fluorescence detector, are presented.

## 1 Introduction

The energy range between $10^{17}$ eV and $4 \times 10^{18}$ eV is of great interest for understanding the origin of cosmic rays. At these energies the transition from the galactic to extragalactic accelerators [1, 2, 3] is expected. Also a spectral feature caused by the drop of the heavy component of the galactic cosmic rays [4] has been predicted. The Pierre Auger Observatory [5], being the largest cosmic ray experiment in operation, has delivered important results for solving the nature of cosmic rays above an energy of $10^{18}$ eV. To extend the measurements to lower energies two enhancements are being built: HEAT [6] (High Elevation Auger Telescopes) and AMIGA (Auger Muons and Infill for the Ground Array) [7]. We present the performance and current status of the analysis of the data taken with the infill array of AMIGA. Construction of the infill array began in 2008: currently 53 stations (87% of the total 61) have been deployed. The detectors are on a triangular grid with a spacing of 750 m. The reconstruction procedures are adaptations of those used for the 3000 km$^2$ array where the spacing between the detectors is 1500 m. In the first part we describe the trigger efficiency and the accumulated exposure, followed by the description of the reconstruction of air-showers.

## 2 Trigger efficiency and acceptance

The trigger system of the infill array is adopted from the regular Auger array. An event is accepted when at least 3 stations forming a triangle satisfy a local trigger of the type Time-over-Threshold (3ToT event) [8]. The smaller spacing between stations of the infill lead to an increase of the trigger efficiency at low energy. The trigger efficiency as a function of energy for 3ToT events with zenith angles below 55° is illustrated in Fig. 1, for both infill and regular array. The calculation is based on the parametrization of the single station lateral trigger probability [9], which reflects the properties of the station response and of the air-shower development. A simple test of the threshold energy where the array is fully efficient is performed via a $\chi^2$ test of an isotropic flux of the observed cosmic rays. Below this energy the trigger depends on the zenith angle. The reduced $\chi^2$ as a function of the energy estimator, $S_{35}$, is illustrated in the inset of Fig. 1. The trigger is independent of the zenith angle for $S_{35}$ larger than $20\,\mathrm{VEM}$, corresponding to an energy of $\approx 3 \times 10^{17}$ eV. From both methods we conclude that above this energy the infill array is 100% efficient for cosmic rays with a zenith angle of less than 55°.

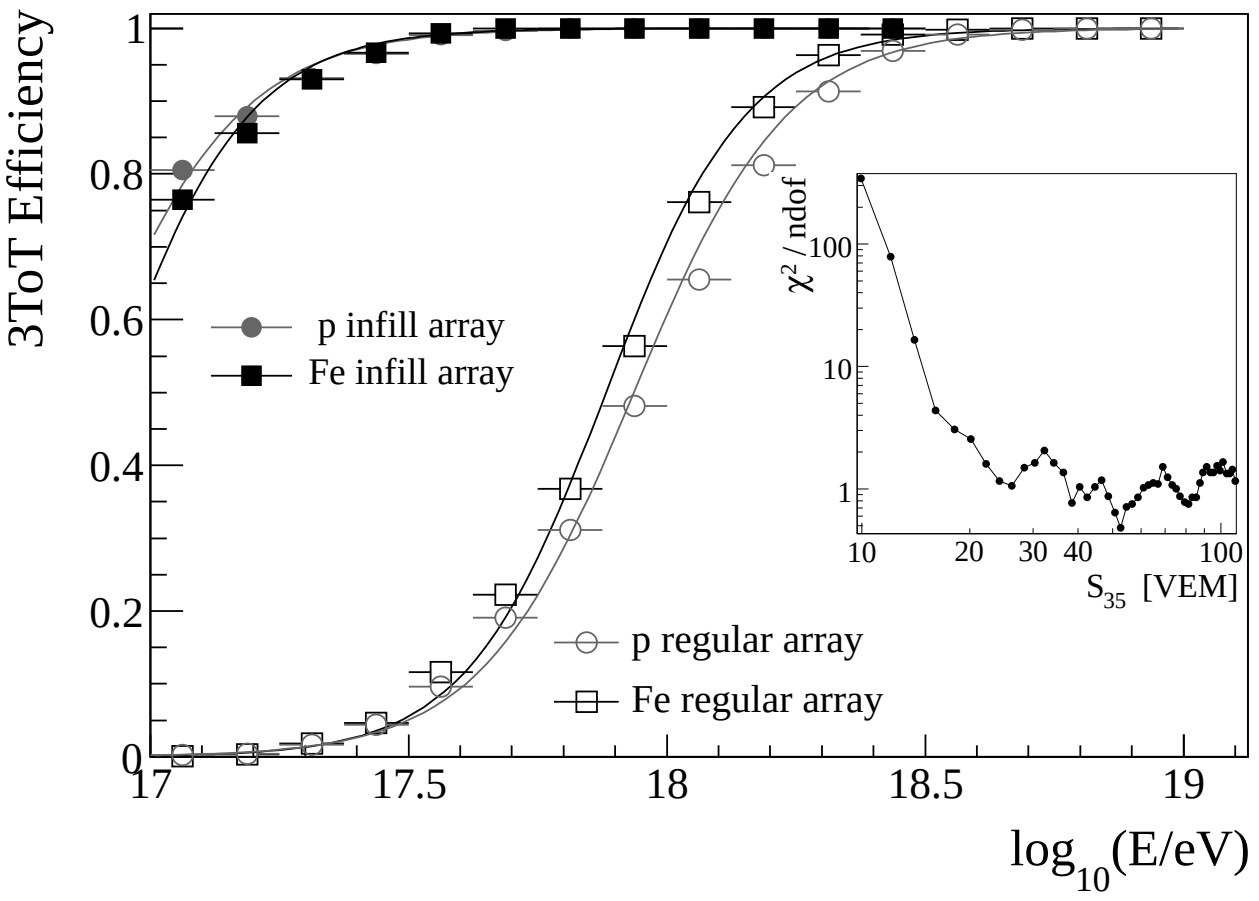

Figure 1: 3ToT trigger efficiency for the infill and regular array obtained from simulations of Iron and proton primaries. In the inset figure the reduced $\chi^2$ for a flat dependency on the zenith angle of the trigger is depicted.

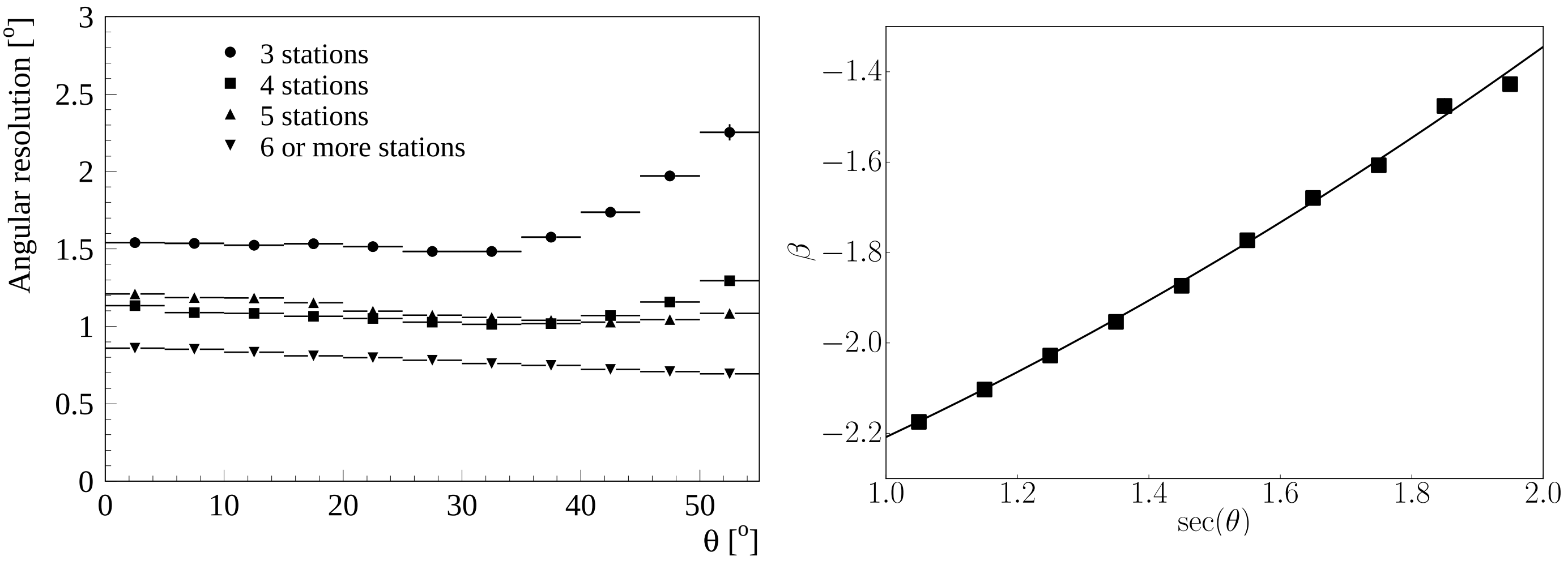

Figure 2: (left) Angular resolution for different multiplicities as a function of the zenith angle. (right) The weighted mean of the LDF parameter, $\beta$, as a function of the zenith angle. The line represents the parametrization of $\beta$.

To guarantee the selection of good quality and well-contained events, a fiducial cut (T5) is applied so that only events in which the station with the highest signal is surrounded by 6 working neighbors (i.e. a working hexagon) are accepted. This condition assures a good reconstruction of the impact point on ground, meanwhile allowing for a simple geometrical calculation of the aperture. The 3ToT trigger rate is $(55 \pm 6)$ events/day/hexagon out of which $(28 \pm 3)$ events/day/hexagon are T5. Integrating the instantaneous effective area over the time when the detector was stable, the acceptance between August 2008 (when the first 3 hexagons of the infill were completed) and March 2011 (16 hexagons) amounts to $(26.4 \pm 1.3)\ \mathrm{km}^2\ \mathrm{sr\ yr}$. With the current configuration we record (390 ± 70) T5 events/day and the data sample contains more than 260,000 T5 events.

## 3 Reconstruction of the air-showers

The reconstruction algorithm for the events triggering the infill array is based on the well-tuned code for the regular surface detector array. After selecting the signals which are generated by air-showers, the direction and the energy of the primary cosmic ray are deduced from the timing information and from the total recorded signal in the stations. The atmospheric muons can generate background signal in a time window close to the arrival of air-shower particles. The stations are selected according to their time compatibility with the estimated shower front. The time cuts were determined such that 99% of the stations containing a physical signal from the shower are kept. The algorithm for the signal search in the time traces rejects further accidental signals by searching for time-compatible peaks.

**Angular resolution:** The arrival direction is obtained from the time propagation of the shower front on the ground which is approximated as a sphere with the origin on the shower axis traveling with the speed of light. To obtain the angular resolution [10] the single station time variance is modeled to take into account the size of the total signal and its time evolution. The angular resolution achieved, illustrated in Fig. 2(left), for events with more than 3 stations is better than 1.3° and is better than 1° for events with more than 6 stations.

**Lateral distribution function:** The impact point on ground of the air showers is deduced in the fit of the lateral distribution of the signals as well. The fit of the lateral distribution function (LDF) [11] is based on a maximum likelihood method which also takes into account the probabilities for the stations that did not trigger and the stations close to the shower axis which are saturated. The saturation is caused by the overflow of the FADC read-out electronics and a modification of the signal occurs due to the transition of the PMTs from a linear to a non-linear behavior. Two functions have been investigated to describe the lateral distribution of the signals on ground: a log-log parabola (LLP), used in the current analysis to infer the systematic uncertainties due to the LDF assumption, and a modified Nishimura-Kamata-Greisen (NKG) function:

$$S(r) = S(r_{\mathrm{opt}}) \left( \frac{r}{r_{\mathrm{opt}}} \frac{r + 700\,\mathrm{m}}{r_{\mathrm{opt}} + 700\,\mathrm{m}} \right)^{\beta} \qquad (1)$$

where $r_{\mathrm{opt}}$ is the optimum distance and $S(r_{\mathrm{opt}})$ is used to obtain an energy estimator. The parameter $\beta$ depends on zenith angle and on energy. For events with only 3 stations the reconstruction of the air-showers can be obtained only by fixing the $\beta$ parameter. To obtain the parametrization of $\beta$, shown in Fig. 2(right), events with more than 6 stations are selected. The vertical events are observed at an earlier shower age than the inclined ones, thus having a steeper LDF due to the different contributions from the muonic and electromagnetic components on ground. The dependence of the LDF parameters on the energy is under investigation.

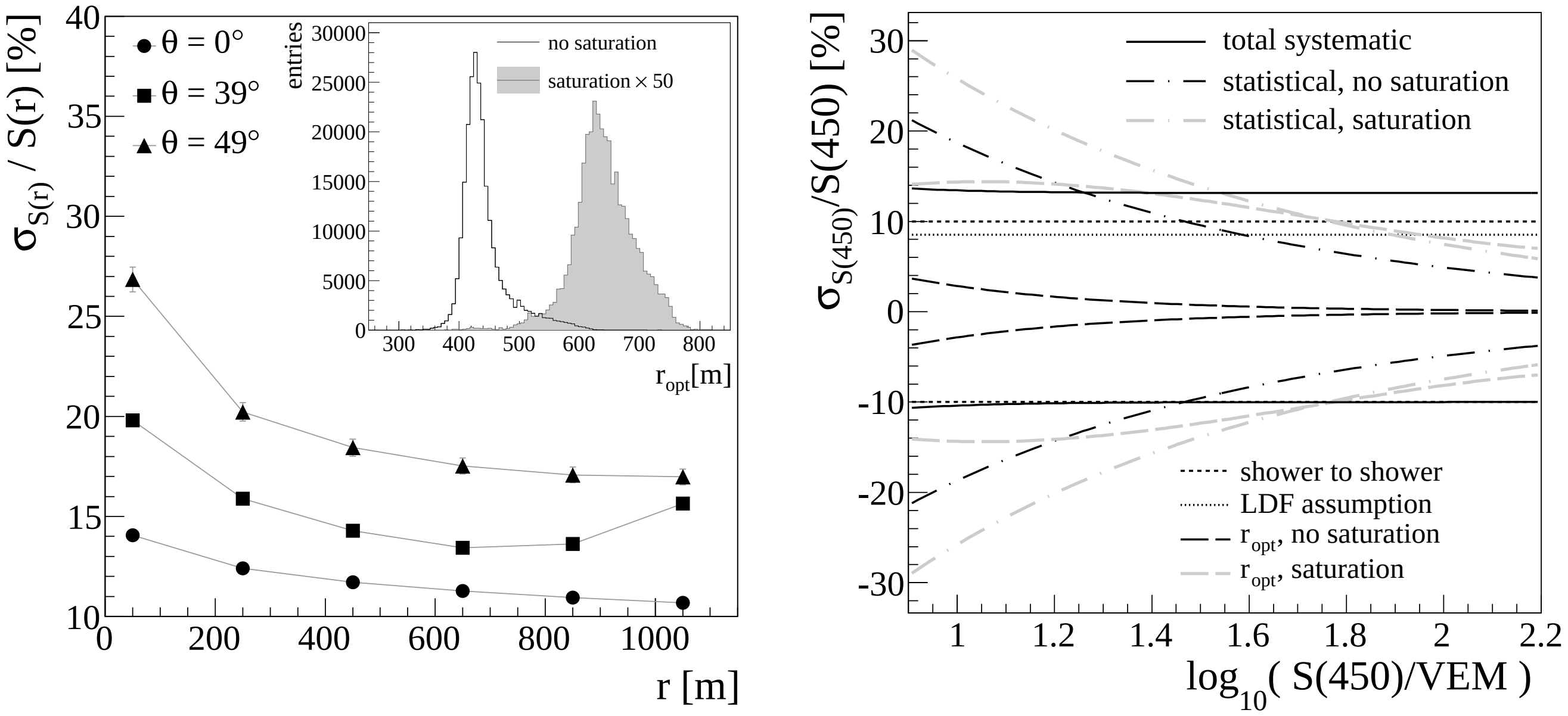

Figure 3: (left) The relative uncertainty of the signal as a function of the distance to the shower axis for different zenith angles deduced from simulations of a mixed composition (50% p, 50% Fe) at $5.62 \times 10^{17}$ eV. The distributions of the optimum distance obtained from data for the events without saturation and for events with a saturated signal (multiplied by 50, gray) are shown in the inset. (right) Relative statistical and systematic uncertainties of S(450). For the description of different contributions see text.

**Optimum distance and uncertainties of S(450):** The optimum distance is defined as the distance from the shower axis where the fluctuations of the LDF are minimized and it is mostly determined by the spacing between the detectors [12]. In Fig. 3(left) the relative uncertainties of the signals at different distances from the shower axis are illustrated for simulations of a mixed composition (50% proton and 50% iron) at an energy of $5.62 \times 10^{17}$ eV and different zenith angles. The air-shower simulations were performed with CORSIKA [13], using QGSJet-II [14] and Fluka [15], and the detector simulation were based on Offline [16]. The uncertainties, containing the shower-to-shower fluctuations, are minimized at distances larger than 450 m.

From data we obtained the distance where the LDF is least sensitive to the $\beta$ parameter by performing the reconstruction of the same event with different $\beta$ values within the uncertainties of the parametrization. This is illustrated in the inset of Fig. 3(left). Similar to the regular array [17], we distinguish the events where the signal in the station closest to the shower axis is saturated. The mean of the distribution of the distance to the shower axis where the impact of the LDF is minimal, for the events without saturated signals is $(442.3 \pm 0.1)\,\mathrm{m}$ with a RMS of $(40.33 \pm 0.06)\,\mathrm{m}$, while for the events with at least a saturated signal is $(639.1 \pm 0.1)\,\mathrm{m}$ with a RMS of $(51.64 \pm 0.06)\,\mathrm{m}$. The signal at $r_{\mathrm{opt}} = 450\,\mathrm{m}$, S(450), was chosen to obtain an energy estimator.

The statistical and systematic uncertainties of S(450) as a function of $\log_{10} S(450)$ are illustrated in Fig. 3(right). The parametrizations were obtained from data similar to the regular array [17]. The statistical uncertainties of S(450) vary from 20 % at 10 VEM to 5 % at 100 VEM. The events in which at least one signal is saturated have an uncertainty that is larger by $\approx 10\%$. The $r_{\mathrm{opt}}$ changes from event to event. The contribution of the variations of $r_{\mathrm{opt}}$ to the uncertainty of S(450) was obtained by reconstructing the same event with different $\beta$ values. While this effect is negligible for the accuracy of S(450) for events without saturation, it contributes with 15% to the total uncertainty for saturated events. The relative difference of the S(450) obtained using a LLP to the estimation from a NKG function is +8.5%. This systematic uncertainty cancels in the final energy resolution via the cross-calibration with the fluorescence detector energy. Preliminary estimations let us assume that the shower-to-shower fluctuations contribute with 10% to the total uncertainty.

**Attenuation in the atmosphere:** S(450) is corrected for the zenith angle dependency caused by the shower attenuation in the atmosphere with a constant intensity cut method [18]. The zenith angle correction has been deduced empirically to be a second degree polynomial in $x = \cos^2\theta - \cos^2 35°$. The zenith angle of 35° represents the median of the distribution of the arrival directions of the observed cosmic rays. The equivalent signal at 35°, $S_{35}$ is used to infer the energy: $S_{35} = S(450)/(1+(1.59\pm$ $0.05)x - (1.14 \pm 0.21)x^2)$. The number of events as a function of $S_{35}$ is illustrated in Fig. 4 for different zenith angle intervals. The trigger threshold effect can be seen below 20 VEM. The $S_{35}$ spectrum extends to more than 2000 VEM, corresponding to an energy of $\approx 3 \times 10^{19}$ eV.

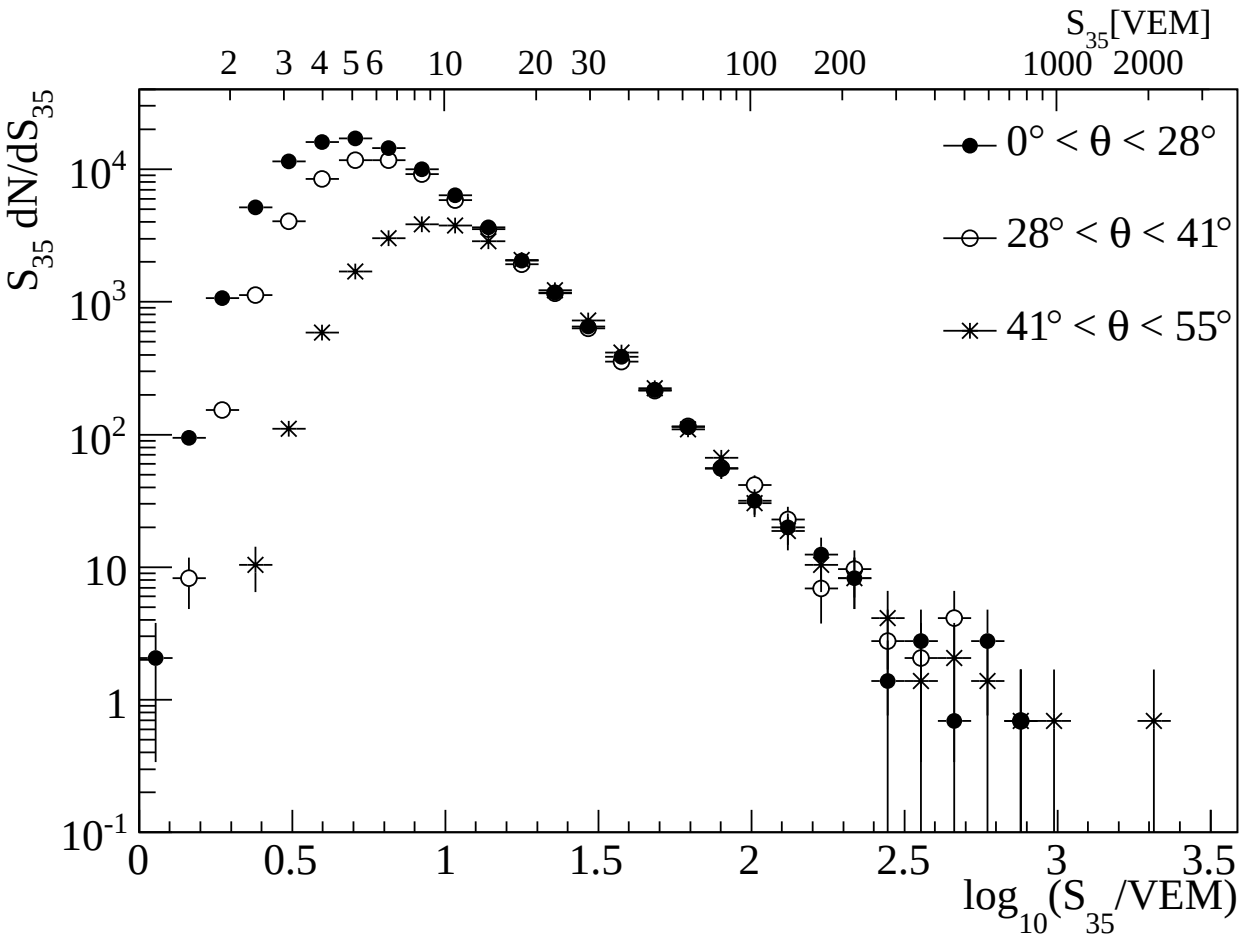

Figure 4: The distribution of events as a function of $\log_{10}(S_{35}/\mathrm{VEM})$ for different zenith angle intervals.

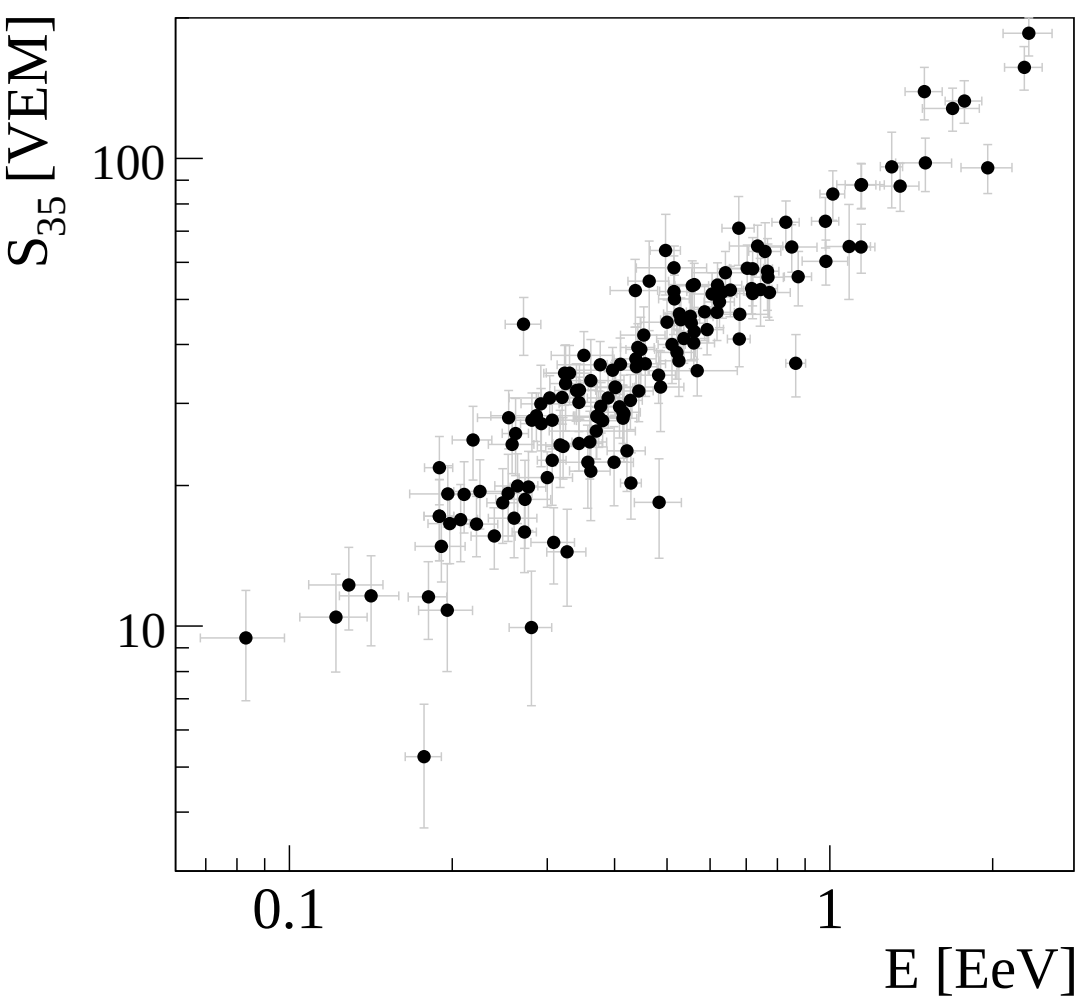

Figure 5: The correlation between $S_{35}$ and energy.

**Energy calibration:** The energy calibration of the surface detector data is obtained from air-showers that were simultaneously measured with the fluorescence detector (i.e. hybrid events). At energies lower than $10^{18}$ eV due to its field of view, the fluorescence detector observes more deep showers than shallow ones. Therefore it is necessary to apply a fiducial field of view cut [19] that ensures an unbiased energy calibration. The selection criteria and the energy systematic uncertainties are currently under study. $S_{35}$ shows a strong correlation with the energy as it is illustrated in Fig. 5 for the selected [18] hybrid events.

## 4 Conclusions

The infill detector, part of the AMIGA experiment, has been operating in good conditions since its deployment. It extends the energy range for the surface detector of Pierre Auger Observatory down to $3 \times 10^{17}$ eV. The analysis, based on the algorithms developed for the regular array is in an advanced stage. The integrated exposure between August 2008 and March 2011 is $(26.4 \pm 1.3)$ km$^2$ sr yr. The achieved angular resolution is better than 1 degree for an energy of $3 \times 10^{17}$ eV. The resolution of S(450) varies from 20% at 10 VEM to 10% at 100 VEM, at the highest energies being dominated by shower-to-shower fluctuations. The studies on the selection of hybrid events used for the energy calibration and on the energy systematic uncertainties are in progress.

# Energy calibration of data recorded with the surface detectors of the Pierre Auger Observatory: an update

ROBERTO PESCE[1] FOR THE PIERRE AUGER COLLABORATION[2]
[1] *Università di Genova and INFN, Via Dodecaneso 33, 16146 Genova, Italy*
[2] *Observatorio Pierre Auger, Av. San Martin Norte 304, 5613 Malargüe, Argentina*
*(Full author list: http://www.auger.org/archive/authors_2011_05.html)*
*auger_spokespersons@fnal.gov*

**Abstract:** The energy of the primary particles of air showers recorded using the water-Cherenkov detectors of the Pierre Auger Observatory is inferred from simultaneous measurements of showers with those detectors together with the fluorescence telescopes. The signal on the ground at 1000 m from the shower axis obtained using the water-Cherenkov detectors is related directly to the calorimetric energy measured with the telescopes. The energy assignment is therefore independent of air shower simulations except for the assumptions that must be made about the energy carried into the ground by neutrinos and muons. The correlation between the signal at the ground and the calorimetric energy is used to derive a calibration curve. Taking advantage of increased statistics with respect to previous publications we present an update and improvement of the method used to determine the energy scale. The systematic uncertainties of the calibration procedure are addressed.

## 1 Introduction

The Pierre Auger Observatory [1] is used to detect extensive air showers (EAS) with an array of over 1600 water-Cherenkov detectors, collectively called the Surface Detector Array (SD). The SD measures the lateral distribution of particles on ground with a duty cycle of almost 100% [2]. The SD is overlooked by the Fluorescence Detector (FD), which consists of 27 fluorescence telescopes at four locations on the periphery of the SD. The FD is only used on clear moonless nights and has a duty cycle of 13% [3]. The FD provides a nearly calorimetric energy measurement, since the fluorescence light is produced in proportion to the energy dissipation by a shower in the atmosphere [4, 5].

The signals recorded by the SD are converted into units of vertical-equivalent muons (VEM). One VEM is defined as the average of the signals produced in the 3 PMTs of a water-Cherenkov detector by a vertical muon that passes centrally through it. The EAS axis is obtained from the arrival time of the first particles in each detector station. The core and the lateral distribution function (LDF) are inferred from a global minimization, taking into account the SD station trigger threshold and the overflow of FADC counts in the detectors near the axis of the EAS. In general the energy of the primary particle is correlated with the signal at a fixed distance from the core of the EAS [6]. In this case, the signal at $1000\,\mathrm{m}$ from the axis, $S(1000)$, corrected for the attenuation in the atmosphere (see Section 2), is used as an energy estimator. At this distance, the fluctuations of the signal, due to an imperfect knowledge of the LDF, are minimized [7].

A measurement of the development profile of the air shower (deposited energy versus slant depth) is possible with EAS viewed with the FD in coincidence with the SD. The first step is the determination of the geometry of the axis of the EAS using directions and timing information from the FD pixels, coupled with the arrival time of the shower at the SD station with the highest signal. The procedure results in an arrival direction resolution of better than $1^\circ$. Next, the light collected in the cameras of FD is transformed into the energy deposited along the axis of the shower [8], by taking into account the fluorescence and Cherenkov light contributions and the attenuation of this light by scattering, including multiple scattering [9]. Care is taken to account for the lateral spread of light on the camera due to the emission of both fluorescence and Cherenkov light. The fluorescence light emission along the track of the EAS is converted into energy deposit by using the absolute fluorescence yield in air in the $337\,\mathrm{nm}$ band of $(5.05 \pm 0.71)$ photons$/\mathrm{MeV}$ of energy deposited [10]. This figure is for dry air at $293\,\mathrm{K}$ and $1013\,\mathrm{hPa}$: the wavelength, temperature, pressure and humidity dependence is accounted for using [11]. Due to the limited field of view of the FD, the longitudinal profile is not recorded in its entirety, so a fit with a Gaisser-Hillas function is employed to obtain the full profile. This energy deposit profile is in-tegrated to yield the calorimetric energy, with a correction of about $9\%$ added to take account of the energy carried by high energy muons and neutrinos. This non-detected energy, that is the *invisible energy*, is accounted for by correcting the calorimetric energy $E_{\rm cal}$, detected by the FD. The factor $f_{\rm inv}$ is determined from simulations to obtain the total shower energy $E_{\rm FD} = (1 + f_{\rm inv})\, E_{\rm cal}$. The invisible energy correction is based on the average for proton and iron showers simulated with the QGSJetII model [12] and amounts to about $9\%$ at $10\,\rm EeV$ for a mixed primary composition [13]. The neutrino and muon production probabilities have energy dependencies due to the meson decay probabilities in the atmosphere so that $f_{\rm inv}$ falls from $\sim 10.5\%$ at $1\,\rm EeV$ to $\sim 8.5\%$ at $100\,\rm EeV$ [13]. The factor $f_{\rm inv}$ depends on the hadronic interaction assumptions and is also subject to shower-to-shower fluctuations [14]. The dependence of the energy scale on the hadronic interaction model is below $4\%$.

The sub-sample of EAS that are recorded by both the FD and the SD, called *golden hybrid events*, is used to relate the energy reconstructed with the FD, $E_{\rm FD}$, to $S(1000)$. The energy scale inferred from this data sample is applied to all showers detected by the SD array.

## 2 Data Analysis

A subset of high-quality golden hybrid events detected between 1 January 2004 and 30 September 2010 is used in this analysis, an update and an improvement with respect to the one presented in [15]. Golden hybrid events are those for which the reconstruction of an energy estimator can be derived independently from both the SD and FD data. In this work only events with reconstructed zenith angles less than $60°$ are used.

A fiducial cut is applied to events recorded by the SD to ensure adequate containment inside the array, and hence a reliable core reconstruction and estimate of $S(1000)$. The cut requires that six active stations surround the station with the highest signal [16]. The water-Cherenkov detector with the highest signal must be within $750\,\rm m$ of the shower axis [17]. The reduced $\chi^2$ of the longitudinal profile fit to the Gaisser-Hillas function has to be less than $2.5$. Furthermore the $\chi^2$ of a linear fit to the longitudinal profile has to exceed the Gaisser-Hillas fit $\chi^2$ by at least $4$ [8]. The depth of shower maximum, $X_{\rm max}$, must be within the field of view of the telescopes and the fraction of the signal detected by the FD and attributed to Cherenkov light must be less than $50\%$. The uncertainties on $E_{\rm FD}$ and on $X_{\rm max}$ are required to be less than $20\%$ and $40\,\rm g/cm^2$ respectively. The selection criteria include a measurement of the vertical aerosol optical depth profile (VAOD) made using laser shots generated by the central laser facility (CLF) [18] and observed by the fluorescence telescopes in the same hour of each selected hybrid event; the VAOD value must be smaller than $0.1$. Furthermore the cloud fraction in the field of view, measured from the information provided by the LIDAR systems of the observatory [18], is required to be less than $25\%$. The limited field of view of the FD and the requirement to observe the EAS maximum may introduce a dependency in the event selection on the primary mass. To avoid this effect, a fiducial cut on the slant depth range observed by the telescopes is performed [19], ensuring that the field of view is large enough to observe all plausible values of $X_{\rm max}$. This cut is introduced for the first time in this analysis, taking advantage of the increased statistics of data. The effect of this fiducial cut is to reject about $22\%$ of events above $3\,\rm EeV$ and $4\%$ above $10\,\rm EeV$. As explained in Section 3, the application of this cut does not change the results of the energy calibration significantly. Applying these cuts, a set of 839 golden hybrid events with energy $E_{\rm FD} \geq 3\,\rm EeV$ (where the SD trigger is fully efficient [16]) is selected.

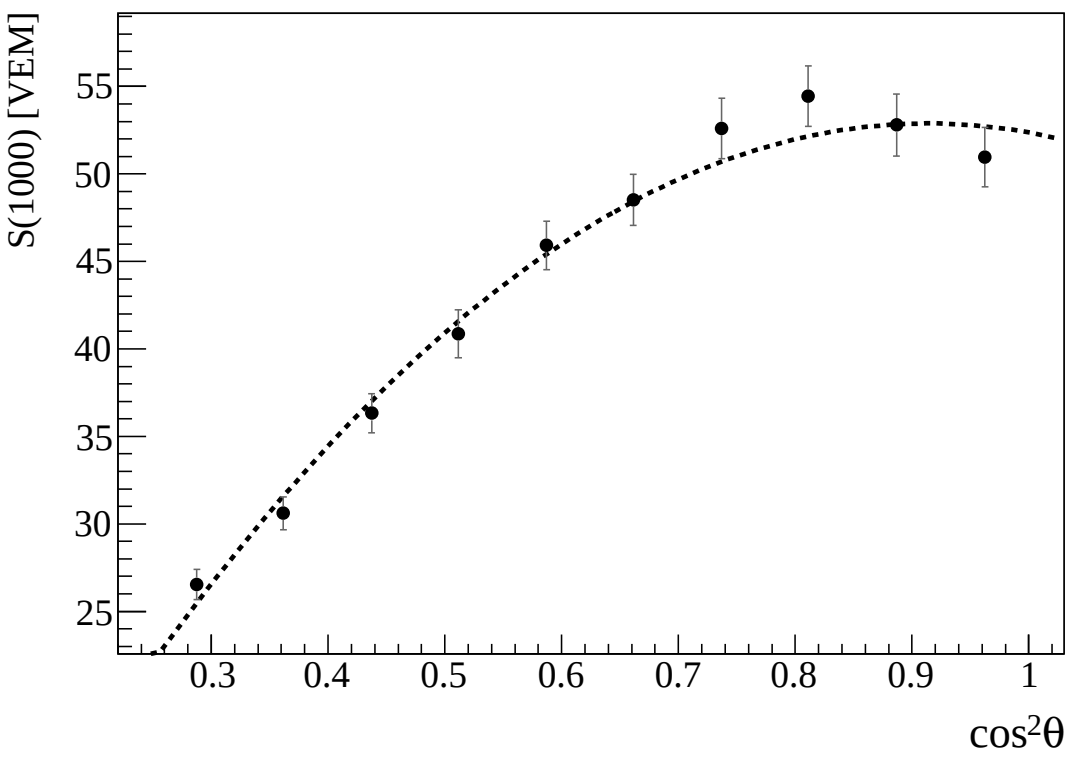

Figure 1: Attenuation curve, $\mathrm{CIC}(\theta)$ fitted with a second degree polynomial in $x = \cos^2\theta - \cos^2\bar{\theta}$.

For a given energy, the value of $S(1000)$ decreases with the zenith angle, $\theta$, due to the attenuation of the shower particles and geometrical effects. Assuming an isotropic flux of primary cosmic rays, we extract the shape of the attenuation curve from the data using the constant intensity cut method [20]. The attenuation curve has been fitted with a second degree polynomial in $x = \cos^2\theta - \cos^2\bar{\theta}$: $\mathrm{CIC}(\theta) = 1 + a\,x + b\,x^2$, where $a = (0.87 \pm 0.04)$ and $b = (-1.49 \pm 0.20)$. The attenuation curve is shown in Fig. 1. The average angle, $\bar{\theta} \simeq 38°$, is taken as a reference to convert $S(1000)$ to $S_{38} \equiv S(1000)/\mathrm{CIC}(\theta)$. $S_{38}$ may be regarded as the signal $S(1000)$ the shower would have produced if it had arrived at $\theta = 38°$. The values of the parameters $a$, $b$ are deduced for $S_{38} = 47$ VEM, that corresponds to an energy of about $9\,\rm EeV$ (see Section 3). The relative difference of the $\mathrm{CIC}(\theta)$ with respect to the previous analysis [15] is about $(-4 \pm 4)\%$ for $\theta = 0°$ and $(-4 \pm 9)\%$ for $\theta = 60°$, i.e. the values of $S_{38}$ are reduced by about $4\%$.

The reconstruction accuracy $\sigma_{S(1000)}$ of $S(1000)$ is composed of three contributions: a statistical uncertainty due to the finite number of particles intercepted by a given SD station and the limited dynamic range of the signal detection; a systematic uncertainty due to assumptions on the shape of the lateral distribution function; and an uncertainty due to shower-to-shower fluctuations [21]. The last term con-tributes a factor of about $10\%$, while the contribution of the first two terms depends on energy and varies from 20% (at $S(1000) = 1.5$ VEM, equivalent to $\sim 0.3\,\mathrm{EeV}$) to $6\%$ (at 200 VEM, equivalent to $\sim 40\,\mathrm{EeV}$).

The FD energy resolution is determined by propagating the statistical uncertainty on the light flux, the invisible energy uncertainty due to shower fluctuations and the uncertainties on EAS geometry and VAOD profiles. The overall energy resolution is $7.6\%$ and it is almost constant with energy.

## 3 Energy Calibration

The analysis of the golden hybrid events leads to a relation between $S_{38}$ and $E_{\mathrm{FD}}$. The main challenge in this part of the analysis is to suppress the bias coming from the inclusion of events with energy below the trigger saturation threshold. The SD is fully efficient above energies of $3\,\mathrm{EeV}$ [16]. The upward fluctuations of $S(1000)$ below this energy would introduce a large bias in the energy conversion. In our past work [15], events below the threshold energy were rejected by a $\chi^2$ method. As an evolution of this procedure, in the present study, a maximum likelihood (ML) method is used (see also [22]). This method, as the previous one, is based only on the data and does not depend on simulations. The ML function takes into account the evolution of uncertainties with energy, as well as event migrations due to the finite resolution of the SD. The ML method has been tested with the dataset used in the previous analysis and reproduces the same results as the $\chi^2$ method: the ML method is mathematically more rigorous. The method has then been applied to the present sample of 839 selected hybrid events with energy $E_{\mathrm{FD}} \geq 3\,\mathrm{EeV}$ (see Section 2).

The relation between $S_{38}$ and $E_{\mathrm{FD}}$ is well described by a single power-law function,

$$E_{\mathrm{FD}} = A\, S_{38}^{B}, \tag{1}$$

where the resulting parameters from the data fit are $A = (1.68 \pm 0.05) \times 10^{17}\,\mathrm{eV}$ and $B = 1.035 \pm 0.009$. The most energetic selected event has an energy of about $75\,\mathrm{EeV}$.

The relative difference in the energy measured by the SD, $E_{\mathrm{SD}}(S_{38})$, using this energy calibration and the previous one [15] is tabulated in the second column of Table 1. The changes in the energy scale are due to an update of the absolute calibration of the FD pixels and improvements to the FD reconstruction, which now properly treat the lateral width of Cherenkov emission, multiple scattering of light, and the temperature and humidity dependence of quenching of fluorescence emission. Part of the difference between this energy calibration and the previous one is due to the introduction of the fiducial cuts. When not applying them, the energy changes by the factor reported in the third column of Table 1. The changes in calibration curves are smaller than the systematic uncertainty due to the application of the calibration method, of about $7\%$ at $10\,\mathrm{EeV}$ and $15\%$ at $100\,\mathrm{EeV}$ [15].

| $E_{\mathrm{FD}}$ | $E_{\mathrm{new}}/E_{\mathrm{old}} - 1$ | $E_{\mathrm{new}}/E_{\mathrm{nofid}} - 1$ |
|---|---|---|
| $3\,\mathrm{EeV}$ | $(+1.0 \pm 1.7)\%$ | $(+0.4 \pm 1.6)\%$ |
| $10\,\mathrm{EeV}$ | $(-3.1 \pm 1.3)\%$ | $(-1.3 \pm 1.7)\%$ |
| $100\,\mathrm{EeV}$ | $(-10 \pm 3)\%$ | $(-4 \pm 4)\%$ |

Table 1: Relative differences in the new energy calibration, $E_{\mathrm{new}}$, for different values of $E_{\mathrm{FD}}$, with respect to the old calibration [15], $E_{\mathrm{old}}$, (second column) and to the case when no fiducial cuts are applied in the event selection ($E_{\mathrm{nofid}}$, third column).

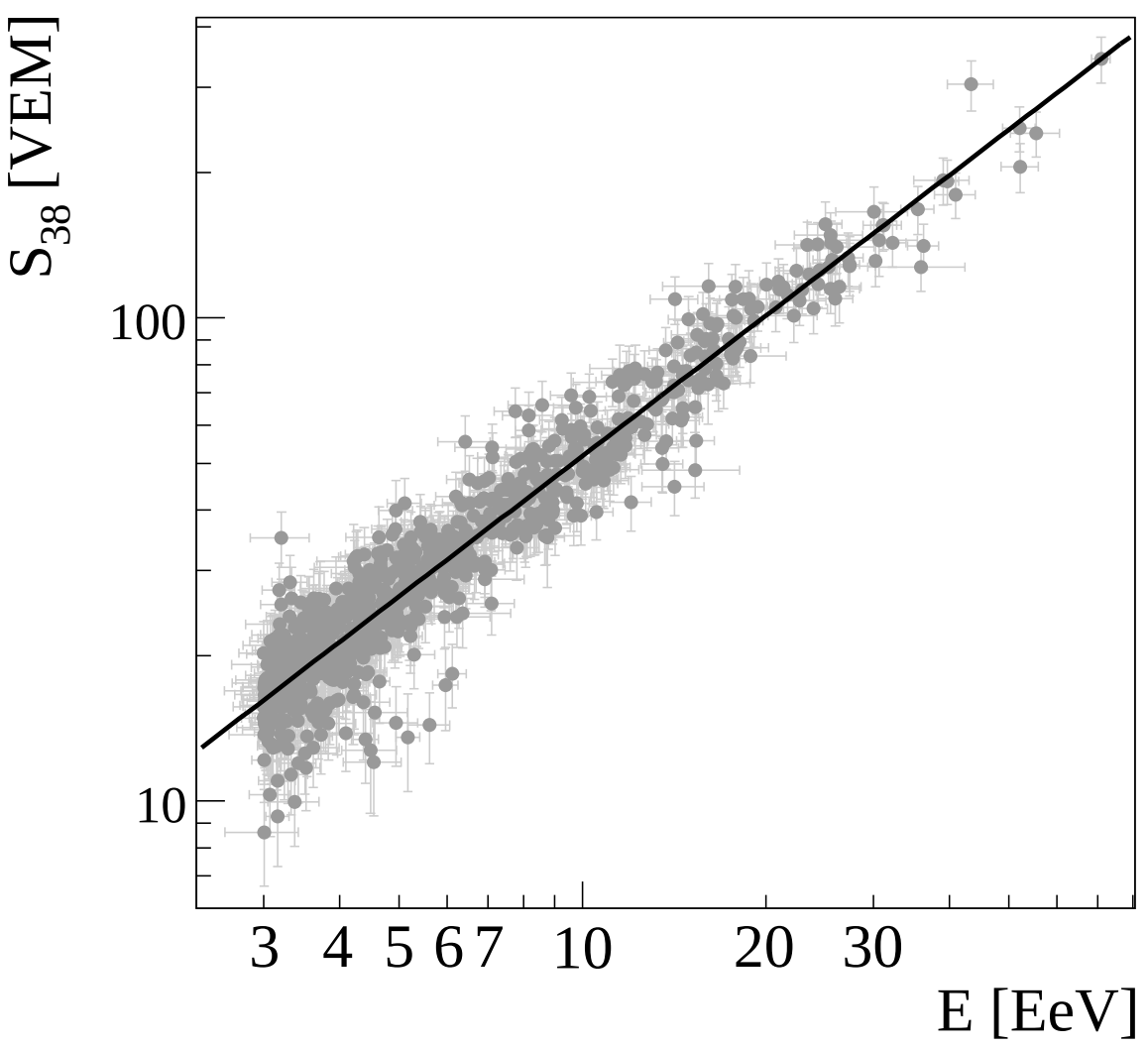

Figure 2: Correlation between $S_{38}$ and $E$ for the 839 selected hybrid events used in the fit. The most energetic event has an energy of about $75\,\mathrm{EeV}$.

The resolution in the SD energy, $E_{\mathrm{SD}}$, can be inferred from the distribution of the ratio $E_{\mathrm{SD}}/E_{\mathrm{FD}}$ [23], fixing the FD energy resolution to the previously quoted $7.6\%$. The fit for three distinct ranges of energy is shown in Figure 3. The resulting SD energy resolution, with its statistical uncertainty, is $\sigma_E/E_{\mathrm{SD}} = (15.8 \pm 0.9)\%$ for $3\,\mathrm{EeV} < E_{\mathrm{SD}} < 6\,\mathrm{EeV}$, $\sigma_E/E_{\mathrm{SD}} = (13.0 \pm 1.0)\%$ for $6\,\mathrm{EeV} < E_{\mathrm{SD}} < 10\,\mathrm{EeV}$ and $\sigma_E/E_{\mathrm{SD}} = (12.0 \pm 1.0)\%$ for $E_{\mathrm{SD}} > 10\,\mathrm{EeV}$.

The total systematic uncertainty on the FD energy scale is about $22\%$. It includes contributions from the absolute fluorescence yield ($14\%$) [10], calibration of the fluorescence telescopes ($9.5\%$), the invisible energy correction ($4\%$) [24], systematics in the reconstruction method used to calculate the shower longitudinal profile (10%), and atmospheric effects ($6\% \div 8\%$) [18]. The atmospheric uncertainties include those related to the measurements of aerosol optical depth ($5\% \div 7.5\%$), phase function ($1\%$) and wavelength dependence ($0.5\%$), the atmosphere variability ($1\%$) [25] and the residual uncertainties on the estimation of pressure, temperature and humidity dependence of the fluorescence yield ($1.5\%$).

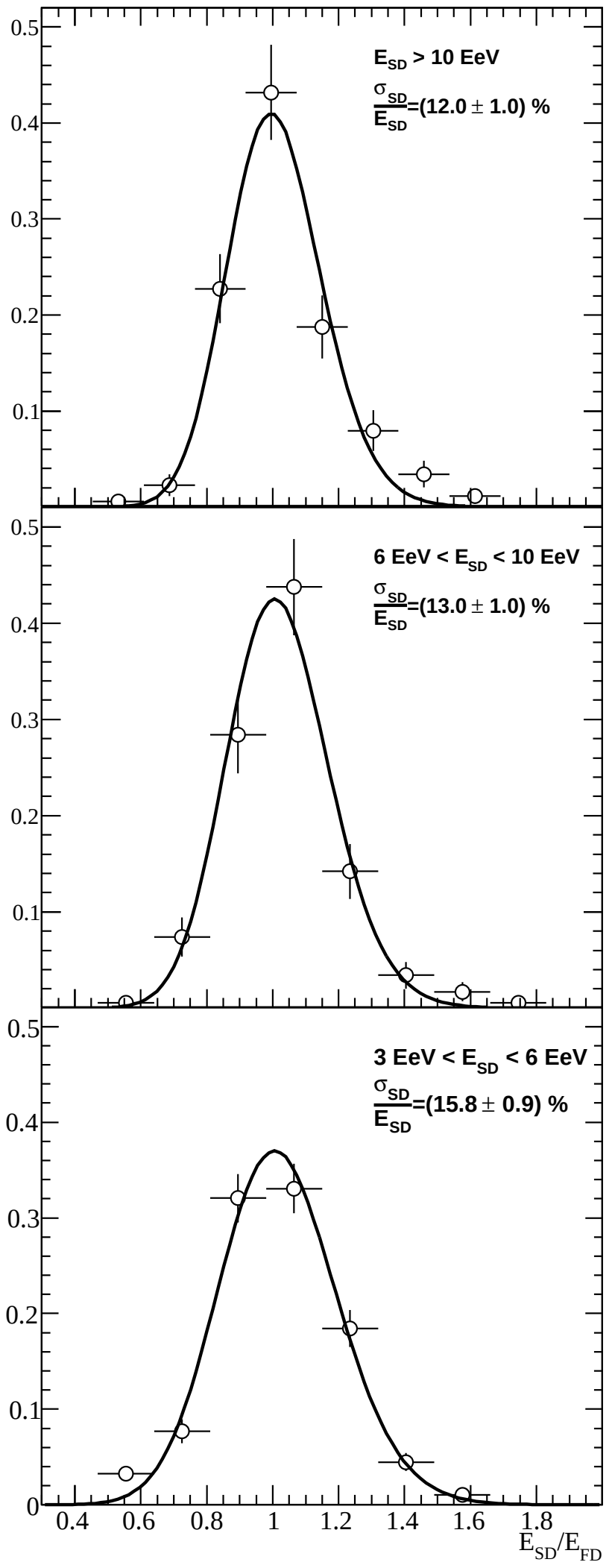

Figure 3: Ratio $E_{\rm SD}/E_{\rm FD}$ for the selected events for various ranges of energy. The lines represent the ratio distribution function for the data [23].

## 4 Conclusions

The energy calibration of the SD array of the Pierre Auger Observatory has been studied using a method based on the data only. It takes into account all the known systematic uncertainties and their dependencies on energy. In this analysis a new fiducial cut in the event selection is also introduced, in order to avoid systematic effects dependent upon the composition of the primary particles.

The results of this method are in good agreement with the previous studies of the energy calibration [15]. The differences are due to some improvements in the energy reconstruction and an update of the calibration constants of the fluorescence telescopes, as well as to the introduction of the new fiducial cut.

The energy spectrum derived from data of the SD array is calibrated using the method presented in this paper and combined with a spectrum based on hybrid data only in [26].

# Reconstruction of inclined showers at the Pierre Auger Observatory: implications for the muon content.

GONZALO RODRIGUEZ[1] FOR THE PIERRE AUGER COLLABORATION[2]
[1] *Departamento de Física de Partículas e IGFAE, Universidade de Santiago de Compostela, Spain.*
[2] *Observatorio Pierre Auger, Av. San Martin Norte 304, 5613 Malargüe, Argentina*
*(Full author list: http://www.auger.org/archive/authors_2011_05.html)*
*auger_spokespersons@fnal.gov*

**Abstract:** The properties of extensive air showers (EAS) induced by cosmic rays with zenith angles up to $80^\circ$ can be measured accurately in the surface detector (SD) of the Pierre Auger Observatory. Using a model of the density pattern of muons, extracted from simulations, the shower size, $N_{19}$, related to the total number of muons in EAS, is estimated from the signals measured in the SD stations. The accuracy of the reconstruction of $N_{19}$ is tested using a large sample of simulated events. The shower size is calibrated using the shower energy measured with the fluorescence detector (FD) in a sub-sample of high-quality hybrid events (i.e. events detected simultaneously by SD and FD). This allows the number of muons versus energy to be measured. We compare the number of muons versus energy as obtained through simulations with that measured in data. We find that none of the current shower models, neither for proton nor for iron primaries, are able to predict as many muons as are observed.

## 1 Introduction

The Pierre Auger Observatory [1] is a hybrid cosmic ray air shower instrument for experiment that uses different techniques to detect extensive air-showers. An array of over 1600 water-Cherenkov detectors that covers an area of 3000 km$^2$ with a 1.5 km triangular grid, samples the signal, $S$, as the air shower arrives at the Earth's surface. The Surface Detector (SD) has a 100% duty cycle. The longitudinal shower development is observed by the Fluorescence Detector (FD) that provides a nearly calorimetric measurement of the shower energy. The duty cycle of the FD is approximately 13% [2]. It is possible to relate the energy measured with the FD to the shower size which can be obtained with the SD, using events that can be reconstructed with both the SD and FD, the *golden hybrid events*. This procedure provides the energy calibration [3].

With the SD detector of the Pierre Auger Observatory we have recorded 5936 events between 1 January 2004 and 31 December 2010, with energy above $4 \times 10^{18}$ eV and zenith angle range $62^\circ < \theta < 80^\circ$. Those events represent the 19% of the total data collected, and a 25% increase over the exposure obtained with events $< 60^\circ$. The analysis of inclined events is important because it increases the sky coverage allowing the study of clustering and anisotropy in a larger region of the sky than it is possible with more vertical events. The dominant particles in these events are muons because most of the electromagnetic component is absorbed in the atmosphere. As the total number of muons depends on primary particle type, inclined showers can be used to study composition. Inclined events are also important because they constitute the background in the search for neutrino-induced showers. In this paper we will describe the procedure used to reconstruct the shower size, $N_{19}$, of inclined showers. We will explain how $N_{19}$ is calibrated using the energy measured by FD in events detected simultaneously by SD and FD. Finally we will compare the behavior of the number of muons versus energy as observed with data to that obtained through Monte Carlo (MC) simulations.

## 2 Reconstruction of size parameter $N_{19}$

The arrival direction of a cosmic ray is reconstructed from the signal arrival times by fitting a shower front model [4, 5]. The angular resolution achieved is better than $1^\circ$ at energies $E > 4 \times 10^{18}$eV. The reconstruction of the shower size requires techniques that are different to those used for more vertical showers. A procedure has been developed in which a set of muon densities at the ground, derived from simulations at different energies and zenith angles, is compared with experimental data. This technique was first used to analyse inclined showers recorded at Haverah Park [6] and has been subsequently adapted for the Auger Observatory [7] which uses water-Cherenkov de-tectors of exactly the same depth. Two reconstruction procedures have been developed independently, referred to as Efit and HasOffline in what follows. Using the two procedures provides an opportunity to test for systematic uncertainties arising from the different reconstruction processes.

The shower core position and the size parameter are obtained through a fit of the expected number of muons at each detector, $n_\mu$, to the measured signal. The expected number of muons can be obtained from:

$$n_\mu = N_{19}\rho_\mu^{19}(x - x_c, y - y_c, \theta, \phi)A_\perp(\theta) \qquad (1)$$

where ($x_c$,$y_c$) are the coordinates of the shower core, $\theta$ and $\phi$ are the zenith and the azimuth of the shower direction, $\rho_\mu^{19}$ is the reference muon distribution (the muon map) corresponding to the inferred arrival direction, $A_\perp(\theta)$ is the detector area projected onto the shower plane. It is shown that the dependence of the shape of $\rho_\mu^{19}$ on energy and primary composition can be approximately factorised out and hence the two dimensional distributions of the muon patterns at ground level depend primarily on zenith and azimuthal angles because of the geomagnetic deviation of muons [8]. $N_{19}$ is thus defined as the ratio of the total number of muons, $N_\mu$, in the shower with respect to the total number of muons at $E$=10 EeV given by the reference distribution, $N_{19} = N_\mu(E,\theta)/N_\mu^{\mathrm{map}}(E = 10\ EeV, \theta)$.

Two sets of reference distribution have been produced for protons at $E$=10 EeV. They are based on two different shower simulation packages and two different hadronic interaction models. One of them has been obtained with a set of histograms in two dimensions based on the model developed in [7], using simulations made with AIRES 2.6.0 [9] with QSGJET01 [10] as the choice for the hadronic interaction. The other set is a continuous parameterizations of the patterns obtained directly from the simulations [11] made with CORSIKA 6.72 [12] and the models QSGJETII [13] and FLUKA [14]. Both approaches are valid out to 4 km from the shower axis in the energy range $10^{18}$ eV to $10^{20}$ eV and zenith angle 60° to 88°. The Efit (HasOffline) packages uses the first (second) set based on QGSJET01 (QGSJETII).

From the measured signal $S$ we obtain the muonic signal $S_\mu$ subtracting the average electromagnetic (EM) component. We calculate the EM component in MC simulations in which we obtain $r_{EM} = S_{EM}/S_\mu$, the ratio of the EM to muonic signal. $r_{EM}$ depends on composition, interaction models and shower energy. We have adopted proton AIRES simulations at $10^{19}$ eV with QGSJET01 as a reference for this work. For inclined showers $r_{EM}$ is largest near the shower axis and its effect practically disappears at zenith angles exceeding about $65°$. Once $r_{EM}$ is parameterised the muonic signal is obtained as $S_\mu = S/(1+r_{EM})$

A maximum likelihood method is used to fit the shower size and the core position. This requires knowledge of the probability density function of the measured signal for all the triggered stations. The probability density $p$ to observe a muonic signal, $S_\mu$, in a given station is obtained from the expected number of muons, $n_\mu$, which depends on its relative position with respect to the core:

$$p(S_\mu, n_\mu, \theta) = \sum_{k=0}^{k=\infty} P_{oisson}(k; n_\mu)P_{st}(k, S_\mu, \theta)P_{Tr}(S_\mu).$$

Here $P_{oisson}(k;n_\mu)$ gives the Poisson probability that $k$ muons go through the detector when $n_\mu$ are expected, $P_{Tr}(S_\mu)$ is the probability of triggering and $P_{st}(k, S_\mu, \theta)$ is the probability density function (p.d.f.) of the muon signal $S_\mu$ for $k$ through-going muons. The response of the SD stations to the passage of a single muon is obtained from a detailed simulation using the standard GEANT4 package [16] within the official software framework of the Observatory [17]. The p.d.f. for $k$ muons, $P_{st}(k, S_\mu, \theta)$, is obtained by convolution. It is important to include details of the non-triggered stations to avoid biases in the reconstruction arising from upward fluctuations of the trigger. These stations provide upper limits to the signals. This can be achieved within the maximum likelihood method, replacing the probability density by the probability that the detectors do not trigger:

$$p(S_\mu = 0, N_\mu, \theta) = \sum_{k=0}^{k=\infty} P_{oisson}(k; N_\mu) \times$$
$$\int_0^\infty dS_\mu(1 - P_{Tr}(S_\mu))P_{st}(k, S_\mu, \theta) \qquad (2)$$

The likelihood function to be maximised is then the product of these probabilities for all stations. In practice only stations that lie within 5 km of the barycenter. The likelihood function depends on three free parameters that have to be fitted, the shower core position ($x_c$,$y_c$) and the shower size $N_{19}$ through the expected number of muons ($N_\mu$) in Eq. (1) which determines the Poisson probabilities. The Poisson distribution takes into account fluctuations in the number of muons entering the detector, while the rest of the fluctuations in the station are introduced through the detector response.

## 3 Accuracy of $N_{19}$ reconstruction

A sample of 100,000 proton showers were generated using AIRES and the hadronic model QGSJET01, with an $E^{-2.6}$ energy spectrum in the range $\log_{10}(E/eV)$=(18.5,20). Showers were chosen from a zenith angle distribution that is flat in $\sin^2\theta$ in the range (50°, 89°), and uniform in azimuthal angle. A further 2700 proton showers were generated using CORSIKA, with the hadronic model QGSJETII and FLUKA. The energy spectrum is flat in $\log_{10}(E/eV)$ in the range (18,20), and the zenith angle has a flat distribution in $\sin^2\theta$ in the range (60°,86°) and a uniform distribution in azimuthal angle. All SD events were generated within the Offline framework [17] with random impact points on the ground.

For simulated showers it is possible to obtain the true value of the size parameter, referred as $N_{19}^{MC}$. In Fig-ure 1 we show the difference between the reconstructed $N_{19}$ and $N_{19}^{MC}$ for both reconstructions. It is plotted as a function of the energy obtained by converting the value of $N_{19}$ into energy according to the calibration data (see figure 3 and related discussions). We note that the average bias for both reconstructions is below ∼4% level in the range $\log_{10}(E/eV)$=(18.5,19.7). We have found agreement between both reconstructions at the 3% level. In the same figure we show the energy resolution, Efit (solid line) and HasOffline (dashed line), going from 25% at $\log_{10}(E/eV) = 18.5$ to 12% at high energies where we expect a more accurate reconstruction as there are more triggered stations.

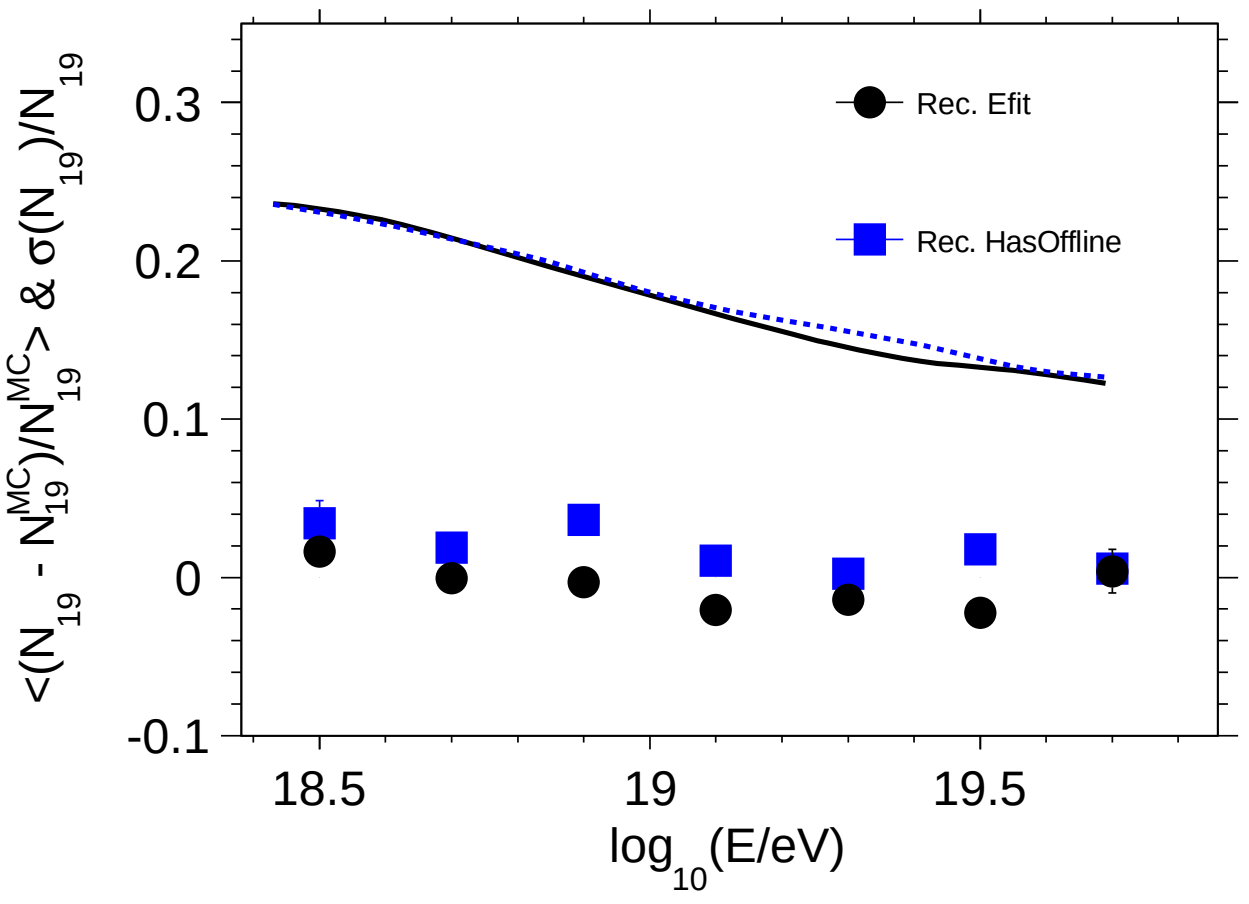

Figure 1: Average(points) and RMS(lines) values of the relative difference between reconstructed ($N_{19}$) and simulated ($N_{19}^{MC}$) shower sizes. Circles and solid line are the results for Efit. Squares and dashed lines for HasOffline.

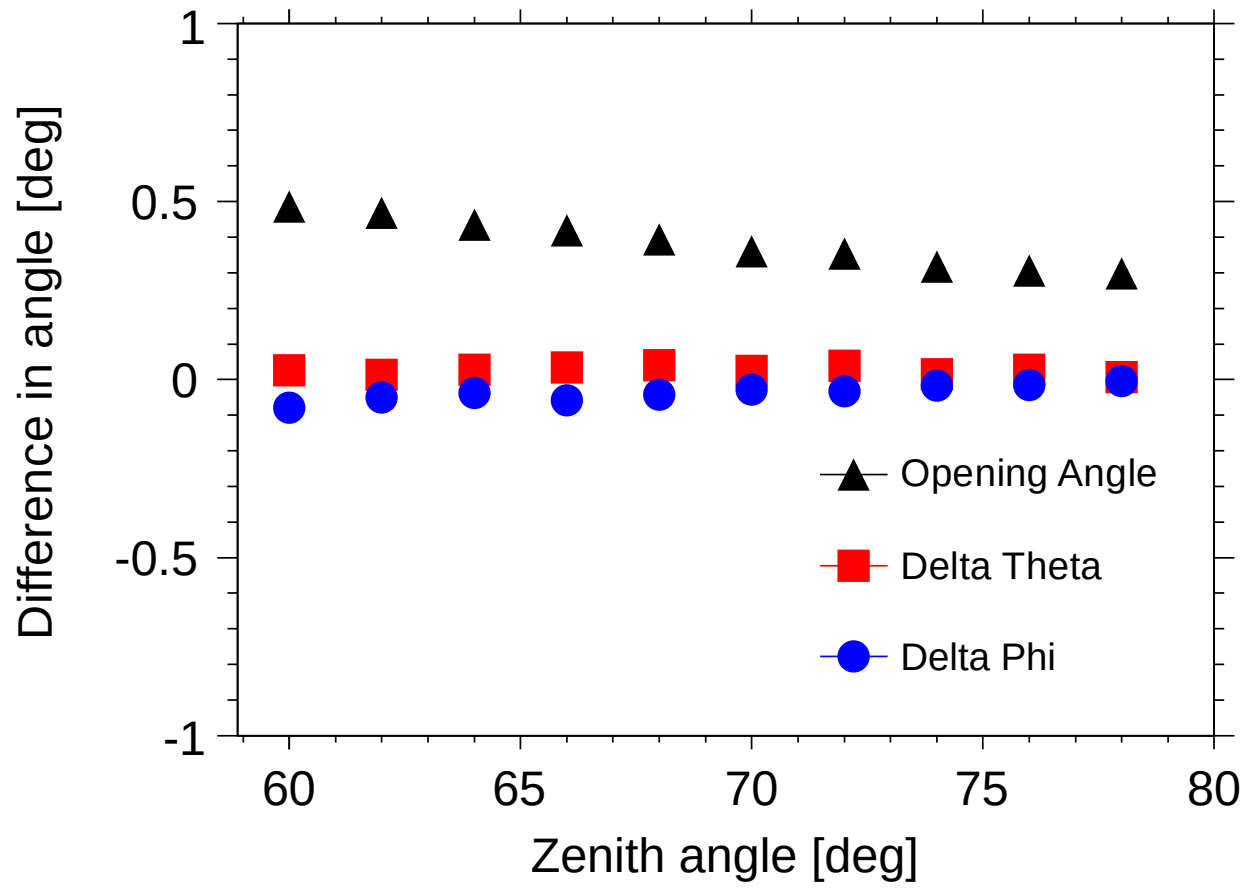

Figure 2: The zenith(square) and azimuth(circles) accuracy as a function of the zenith angle input from MC. The triangles shows the opening angle resolution.

In Figure 2 we show the accuracy in the angular reconstruction as a function of the zenith angle input to the MC calculation. The zenith and azimuth angles are reconstructed with a bias of less than 0.05°, and the opening angle, the angle between the MC and the reconstructed directions, is always less than 0.5°.

## 4 $N_{19}$ versus energy: data versus simulations

The energy of each event is obtained by calibrating $N_{19}$ with *the golden hybrid data set.* In addition the calibration procedure can be used to obtain the number of muons as a function of energy which is sensitive to cosmic ray composition and to the hadronic interactions in the shower. A fit is done using a power law $A(E_{FD}/10\ EeV)^B$ for energies above $4 \times 10^{18}$ eV where the array is 100% efficient. This considerably reduces the systematic uncertainties as the only use of hadronic models comes through the estimate of the energy carried by muons and neutrinos into the ground, the missing or invisible energy. The results of the fit are shown in Fig. 3. From the fit we obtain $A$ = (2.13 ± 0.04 ± 0.11 (sys.)) and $B$ = (0.95 ± 0.02 ± 0.03 (sys.)). Details of the calibration procedure and the data are given respectively in [18] and [3]. A spectrum has been obtained with these data [18].

It is possible to relate $N_{19}$ and shower energy using MC simulation in an analogous way leading to $A_{MC}$ and $B_{MC}$. The results depend on the choices made for composition and hadronic model. In Figure 3 we also show the extreme cases for protons with QGSJETII and for iron nuclei with EPOS1.99 [19].

Using the formula for the calibration fit [18], it is also possible to derive the muon number in the data compared to the predictions from the simulation $N_{19}^{data}/N_{19}^{MC}$ as a function of $E_{FD}$. Simple calculations yield the following expression:

$$\frac{N_{19}^{data}}{N_{19}^{MC}} = \frac{A(E_{FD}/10\ EeV)^B}{A_{MC}(E_{FD}/10\ EeV)^{B_{MC}}} \tag{3}$$

We show the observed number of muons in data compared with the number of muons obtained with the two extreme predictions for proton QGSJETII (solid line) and iron EPOS1.99 (dashed line) in Fig. 4. The grey bands corresponds to the $N_{19}$ systematic uncertainties obtained by comparison between reconstructed and the true values of $N_{19}$ using simulated data. The value of $A_{MC}$ for proton (iron) QGSJETII (EPOS1.99) is 1 (1.7) and that of $B_{MC}$ is 0.934 (0.928). The number of muons deduced from data exceeds that of proton QGSJETII simulations by a factor of 2.1 and that of iron EPOS1.99 by $23\%$. No significant dependence on the energy is obtained in either case.

## 5 Summary and discussion

The reconstruction method for inclined showers has been tested with simulated data. It has been shown that the ave-rage opening angle between the true and reconstructed directions is below $0.5°$ and that the average shower size obtained in the reconstruction reproduces the simulated one within $4\%$. We have shown that the reconstruction of inclined events can be used to extract the number of muons from the data. This is done through the energy calibration that relates the FD energy to $N_{19}$, the muon size with respect to a reference model. When compared to protons simulated with QGSJETII the ratio of the total number of muons at $E_{FD}$=10 EeV is measured to be (2.13 $\pm$ 0.04 $\pm$ 0.11 (sys.)). The $22\%$ systematic uncertainty in the FD energy measurement [2] has not been included.

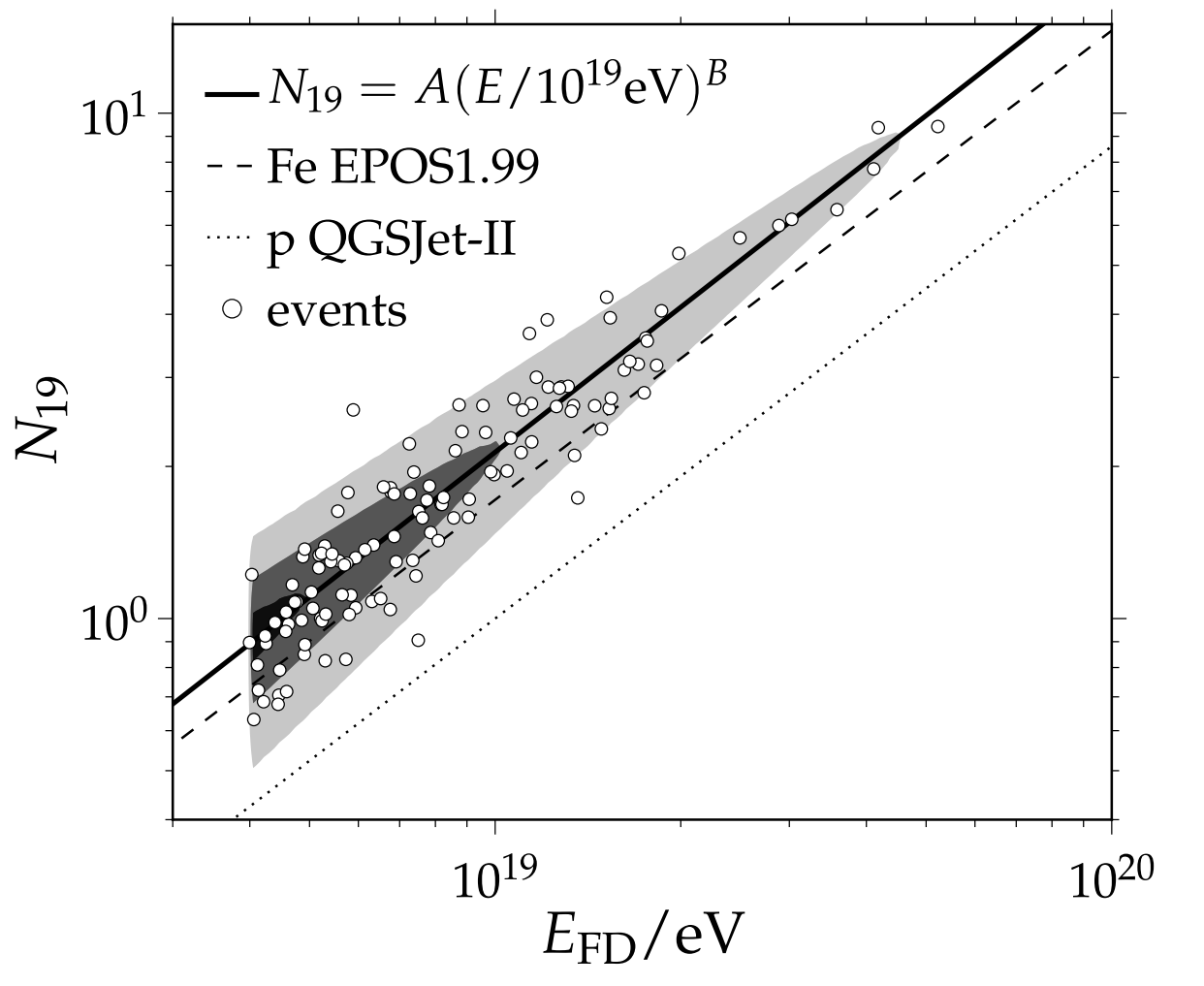

Figure 3: Fit of the calibration curve $N_{19} = A(E/10\ EeV)^B$. The constants $A$ and $B$ are obtained using the maximum-likelihood method. The contours indicate the constant levels of the p.d.f. fhyb integrated over zenith angle, corresponding to 10, 50 and 90% of the maximum value [18]. Calibration curves for proton QGSJETII (dot line) and iron EPOS1.99 (dashed line) are shown for comparison.

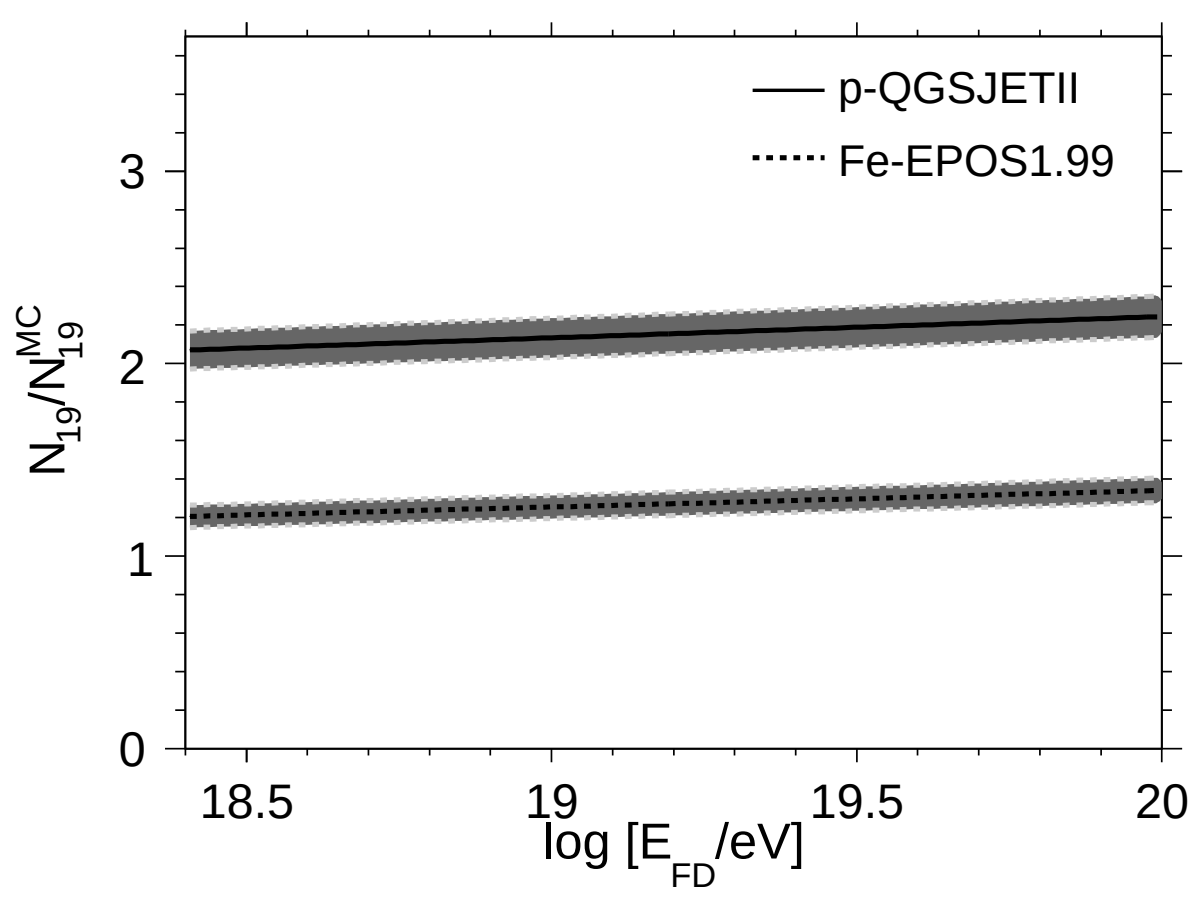

Figure 4: Ratio of the number of muons in data compared to proton QGSJETII (solid line) and iron EPOS1.99 (dashed line) as a function of the energy. The grey bands indicates the systematic uncertanties in $N_{19}$. See text for details.

Several other methods have been developed to obtain the number of muons from the data collected at the Pierre Auger Observatory. All of them use events below $60°$. These methods report similar enhancements of the muon content in the showers [20].

We find that none of the current shower models, neither for proton nor for iron primaries, are able to predict as many muons as are observed.

# Acknowledgments

The successful installation, commissioning and operation of the Pierre Auger Observatory would not have been possible without the strong commitment and effort from the technical and administrative staff in Malargüe.

We are very grateful to the following agencies and organizations for financial support:

Comisión Nacional de Energía Atómica, Fundación Antorchas, Gobierno De La Provincia de Mendoza, Municipalidad de Malargüe, NDM Holdings and Valle Las Leñas, in gratitude for their continuing cooperation over land access, Argentina; the Australian Research Council; Conselho Nacional de Desenvolvimento Científico e Tecnológico (CNPq), Financiadora de Estudos e Projetos (FINEP), Fundação de Amparo à Pesquisa do Estado de Rio de Janeiro (FAPERJ), Fundação de Amparo à Pesquisa do Estado de São Paulo (FAPESP), Ministério de Ciência e Tecnologia (MCT), Brazil; AVCR AV0Z10100502 and AV0Z10100522, GAAV KJB100100904, MSMT-CR LA08016, LC527, 1M06002, and MSM0021620859, Czech Republic; Centre de Calcul IN2P3/CNRS, Centre National de la Recherche Scientifique (CNRS), Conseil Régional Ile-de-France, Département Physique Nucléaire et Corpusculaire (PNC-IN2P3/CNRS), Département Sciences de l’Univers (SDU-INSU/CNRS), France; Bundesministerium für Bildung und Forschung (BMBF), Deutsche Forschungsgemeinschaft (DFG), Finanzministerium Baden-Württemberg, Helmholtz-Gemeinschaft Deutscher Forschungszentren (HGF), Ministerium für Wissenschaft und Forschung, Nordrhein-Westfalen, Ministerium für Wissenschaft, Forschung und Kunst, Baden-Württemberg, Germany; Istituto Nazionale di Fisica Nucleare (INFN), Ministero dell’Istruzione, dell’Università e della Ricerca (MIUR), Italy; Consejo Nacional de Ciencia y Tecnología (CONACYT), Mexico; Ministerie van Onderwijs, Cultuur en Wetenschap, Nederlandse Organisatie voor Wetenschappelijk Onderzoek (NWO), Stichting voor Fundamenteel Onderzoek der Materie (FOM), Netherlands; Ministry of Science and Higher Education, Grant Nos. 1 P03 D 014 30, N202 090 31/0623, and PAP/218/2006, Poland; Fundação para a Ciência e a Tecnologia, Portugal; Ministry for Higher Education, Science, and Technology, Slovenian Research Agency, Slovenia; Comunidad de Madrid, Consejería de Educación de la Comunidad de Castilla La Mancha, FEDER funds, Ministerio de Ciencia e Innovación and Consolider-Ingenio 2010 (CPAN), Xunta de Galicia, Spain; Science and Technology Facilities Council, United Kingdom; Department of Energy, Contract Nos. DE-AC02-07CH11359, DE-FR02-04ER41300, National Science Foundation, Grant No. 0450696, The Grainger Foundation USA; ALFA-EC / HELEN, European Union 6th Framework Program, Grant No. MEIF-CT-2005-025057, European Union 7th Framework Program, Grant No. PIEF-GA-2008-220240, and UNESCO.